\documentclass{article}
\usepackage[utf8]{inputenc} % allow utf-8 input
\usepackage[T1]{fontenc}
\usepackage{microtype}
\usepackage{graphicx}
\usepackage{booktabs} % for professional tables
\usepackage{amsmath, amsfonts, bm, amsthm}
\usepackage{enumerate}
\usepackage{hyperref}
\usepackage{epstopdf}
\usepackage{changepage}
\usepackage{algorithm,algorithmic,url}
\usepackage{etoolbox}
\usepackage{setspace,color}
\graphicspath{{./Figures/}}
\usepackage{subcaption}
\theoremstyle{definition}

\usepackage{geometry}
\def\bfe{{\mathbf{e}}}

\def\bfs{{\mathbf{s}}}

\def\bfx{{\mathbf{x}}}
\def\bfy{{\mathbf{y}}}

\def\bfG{{\mathbf{G}}}

\def\bfR{{\mathbf{R}}}
\def\bfS{{\mathbf{S}}}

\def\bfU{{\mathbf{U}}}

\def\bfSigma{{\mathbf{\Sigma}}}

\def \bfalpha {\boldsymbol{\alpha}}

\def \bfeta {\boldsymbol{\eta}}

\def \bfgamma {\boldsymbol{\gamma}}
\def \bfGamma {\boldsymbol{\Gamma}}

\def \bfSigma {\boldsymbol{\Sigma}}
\def \bfeta {\boldsymbol{\eta}}

\usepackage{booktabs}
\def\wh{\widehat}
\def\wt{\widetilde}

\def\N{\mbox{ $\mathcal{N}$}}

\def\var{\hbox{var}}
\def\defby{\stackrel{\mbox{\textrm{\tiny def}}}{=}}

\def\bse{\begin{eqnarray*}}
	\def\ese{\end{eqnarray*}}
\def\be{\begin{eqnarray}}
\def\ee{\end{eqnarray}}
\def\bsq{\begin{equation*}}
\def\esq{\end{equation*}}
\def\bq{\begin{equation}}
\def\eq{\end{equation}}

\title{Accounting for correlated horizontal pleiotropy in two-sample Mendelian randomization using correlated instrumental variants}
\author{Qing Cheng \thanks{Centre for Quantitative Medicine, Duke-NUS Medical School, Singapore,  qing.cheng@duke-nus.edu.sg}
\and Baoluo Sun \thanks{Department of Statistics and Applied Probability, NUS, Singapore, stasb@nus.edu.sg}
	\and Yingcun Xia \thanks{Department of Statistics and Applied Probability, NUS, Singapore, staxyc@nus.edu.sg}
\and Jin Liu \thanks{Centre for Quantitative Medicine, Duke-NUS Medical School, Singapore, {jin.liu@duke-nus.edu.sg}}
}

\date{}
\begin{document}

\maketitle

\begin{abstract}
Mendelian randomization (MR) is a powerful approach to examine the causal relationships between health risk factors and outcomes from observational studies. Due to the proliferation of genome-wide association studies (GWASs) and abundant fully accessible GWASs summary statistics, a variety of two-sample MR methods for summary data have been developed to either detect or account for horizontal pleiotropy, primarily based on the assumption that %the instrument strength independent of direct effect (InSIDE) condition that  
the effects of  variants on exposure ($\bfgamma$) and horizontal pleiotropy ($\bfalpha$) are independent. This assumption is too strict and can be easily violated because of the correlated horizontal pleiotropy (CHP). %, requiring the independence between the effects of  % variants on exposure ($\bfgamma$) and horizontal pleiotropy ($\bfalpha$). 
%However, in many cases, we observe heteroscedasticity in a linear regression for the observed associations between exposures and outcomes. This heteroscedasticity is essentially caused by the correlations between $\bfgamma$ and $\bfalpha$, %horizontal pleiotropy and effects of genetic variants on exposure,
%which is referred to as correlated horizontal pleiotropy (CHP).  %However, accumulating evidence suggests that genetic correlation exists among complex human traits.  
To account for this CHP, we propose a Bayesian approach, MR-Corr$^2$,  that uses the orthogonal projection to reparameterize the bivariate normal  distribution for $\bfgamma$ and $\bfalpha$, %the effects of genetic variants on exposure and horizontal pleiotropy, 
and a spike-slab prior to mitigate the impact of CHP. %The proposed strategy not only can be used to account for CHP in MR, but also applied in the case that there exist associations between genetic variants and unobserved confounders. 
We develop an efficient algorithm with paralleled Gibbs sampling. % providing a statistical insight on how to account for non-zero correlation assumption in the literature of instrumental variable analysis. 
To demonstrate the advantages of MR-Corr$^2$ over existing methods,  we conducted comprehensive simulation studies to compare for both type-I error control and point estimates in various scenarios. By applying MR-Corr$^2$ to study the relationships between pairs in two sets of complex traits, we did not identify the contradictory causal relationship between HDL-c and CAD. Moreover, the results provide a new perspective of the causal network among complex traits. %invalidate the previously arguably causal relationship between HDL-c and CAD. Moreover, the results provide a perspective of causal network among complex traits.  %More over, we further apply MR-Corr$^2$ to  study  the relationships between plasma protein and complex traits and find that. MR-Corr$^2$ invalidates the previous debated causal relationship between HDL-c and CAD and identifies multiple health outcomes, i.e., T2D, CAD, TG, and HDL-c, affected by obese risk factors (BMI and BFP) causally.  
The developed R package and code to reproduce all the results are available at \url{ https://github.com/QingCheng0218/MR.Corr2.}
\end{abstract}

\noindent%
\vfill

\section{Introduction}
\label{sec:intro}

Inferring causal relationships from observational studies is particularly challenging because of unmeasured confounding, reverse causation and selection bias~\cite{davey2003mendelian}. Without adjusting for confounding effects, the relationships obtained from epidemiological studies might be pure associations due to common confounders between the health risk factors (exposures) and outcomes. Conventionally, associations established from randomized controlled trial (RCT) can be taken to be causal as the relationship between potential confounders and health risk factors is broken by randomization. Alternatively, Mendelian randomization (MR) is a study design that can be used to examine the causal effects between exposures and outcomes from observation studies, by mitigating the impact from the unobserved confounding factors~\cite{davey2003mendelian}. As germline genetic variants (single nucleotide polymorphisms, SNPs) are fixed after random mating and independent of subsequent factors, e.g., environment factors and living styles, MR can be taken as a special case of instrumental variable (IV) methods~\cite{didelez2007mendelian}, which has a long history in econometrics and statistics. 
%In parallel with the proliferation of genome-wide association studies (GWASs) and abundant publicly accessible GWAS results, i.e., GWAS summary statistics, MR methods are rapidly gaining popularity. %Similar to IV methods, accounting for horizontal pleiotropy or direct effects is an essential component to make a reliable causal inference. On the other hand, MR methods need to incorporate the consideration for linkage disequilibrium that is a key feature of GWASs.

As a category of methods closely related to IV methods, MR methods require certain assumptions to hold, as shown in Figure~\ref{fig:dag}, to infer the causal relationships between the exposures and outcomes. To relax the {\sl exclusion} assumption (IV3 in Figure~\ref{fig:dag}), a variety of methods have been developed either to detect horizontal pleiotropy, e.g., Q test~\cite{greco2015detecting}, modified Q test~\cite{bowden2019improving}, GSMR~\cite{zhu2018causal},  and MR-PRESSO~\cite{verbanck2018detection}, or to account for horizontal pleiotropy in a joint model, e.g., MR-Egger~\cite{bowden2015mendelian}, MMR~\cite{burgess2015multivariable}, sisVIVE~\cite{kang2016instrumental}, RAPS~\cite{zhao2018statistical},  MRMix~\cite{qi2019mendelian}, BWMR~\cite{zhao2020bayesian} and MR GENIUS~\cite{tchetgen2017genius}. Most developed methods here rely on the instrument strength independent of direct effect (InSIDE) condition~\cite{bowden2017framework}, which requires the independence between the effects of genetic variants on exposure and horizontal pleiotropy. %methods have been developed to relax Assumption X depending on the instrument strength independent of direct effect (InSIDE) condition~\cite{bowden2017framework}, which requires the independence between the effects of genetic variants on exposure and horizontal pleiotropy. 
This crucial assumption for zero correlation was first identified in econometrics literature for IV analysis with direct effects~\cite{kolesar2015identification} and later applied in various MR analysis~\cite{zhao2018statistical, cheng2020mr}.  However, in practice, this assumption is too strict and can be easily violated. %For example, accumulating evidence suggests that genetic correlation exists among complex human traits~\cite{bulik2015atlas}. 
Our motivating example below shows that there exists heteroscedasticity in the linear relationships between many exposure and outcome pairs, indicating the substantial amount of correlation between the effects of genetic variants on exposure and horizontal pleiotropy. %Further studies show that there exists heteroscedasticity in the linear relationships between many exposure and outcome pairs~\cite{morrison2020mendelian}, indicating the substantial amount of correlation between the effects of genetic variants on exposure and horizontal pleiotropy. 
Without correcting for this correlation, MR methods can lead to biased estimates and inflated false-positive causal relationships~\cite{morrison2020mendelian}. %This correlated pleiotropy could be due to the violation of Assumption X and there exist associations between genetic variants and unobserved confounders~\cite{morrison2020mendelian}.  
Recently, \cite{morrison2020mendelian} proposed a new MR method, Causal Analysis Using Summary Effect estimates (CAUSE), %proposed a CAUSE method 
to infer causal effects by removing correlated pleiotropy, that is genetic variants affecting both the exposure and outcome through the unobserved confounders. %However, CAUSE uses step-wise workflow and therefore does not account for uncertainty between steps (More discussion). Moreover, 
Our numerical studies show that CAUSE suffers from severe $p$-value deflation. Statistically rigorous methods are needed to address the problem of heteroscedasticity due to correlated horizontal pleiotropy (CHP) in MR.

On the other hand, most of the classical MR methods are regression-based, e.g., inverse variance weighting (IVW)~\cite{burgess2013mendelian} and MR-Egger~\cite{bowden2015mendelian}, and only work for independent genetic variants. As linkage disequilibrium (LD) is a key feature in GWASs, %essentially a key feature in GWASs, 
a few methods have been developed to account for correlations among  instruments, e.g.,  GSMR~\cite{zhu2018causal}, and MR-LDP~\cite{cheng2020mr}. %, and F-LIML~\cite{patel2020inference}. 
Among them, GSMR can only account for weak or moderate correlations among instruments while MR-LDP can deal with weak instruments from strong correlations. However, none of these methods adjust for CHP. %To further account for correlated horizontal pleiotropy, an MLR strategy can be applied. However, due to the correlations between weak instruments, a nonzero $\alpha_k$ on the $k$-th variant exhibits  nonzero horizontal pleiotropy $\alpha_{k'}(\neq 0)$ nearby the $k$-th variant. Fortunately, LD along the genome can be partitioned into independent blocks and thus horizontal pleiotropy $\alpha_k$ exhibits block sparsity along the genome. 

This paper aims to resolve the issue of correcting for CHP using correlated genetic variants by developing Bayesian models using GWAS summary statistics. In the following, we will briefly introduce the background of MR study, show the problems of MR using a motivating example, discuss the challenges to correct CHP in MR and the need to incorporate weak instruments, and conclude the introduction by outlining our solution.

\subsection{Two-sample MR using GWAS summary statistics}\label{intro.sect1}
As shown in Figure~\ref{fig:dag}, we are interested in inferring causality between an exposure $X$ and an outcome $Y$, where $X$ and $Y$ are confounded by unobserved confounding factors. For the ease of presentation, we assume that genetic variants, $G_1$, $G_2$, $\dots$, $G_p$, are independent here and refer Section~\ref{model.mrcorr2} for details using correlated genetic variants. The corresponding GWAS summary statistics for SNP-exposure and SNP-outcome are denoted as $\{\wh \gamma_k, \wh \bfs_{\gamma_k}^2 \}$ and $\{\wh \Gamma_k, \wh \bfs_{\Gamma_k}^2 \}$,  $\forall k = 1, \dots, p$, usually by performing a simple linear or logistic regression of either exposure $X$ or outcome $Y$ on each of the genetic variants. Assuming that independent samples are used for SNP-exposure and SNP-outcome, for each variant $k$,  we have 
\be 
\label{ssdist1}
\wh \gamma_k \sim \mathcal{N}(\gamma_k,\wh\bfs_{\gamma_k}^2), \wh \Gamma_k \sim \mathcal{N}(\Gamma_k,\wh\bfs_{\Gamma_k}^2), 
\ee
where $\gamma_k$ and $\Gamma_k$ are underlying true coefficients for SNP-exposure and SNP-outcome on variant $k$, respectively. Following the linear structural model (\ref{lsm2}) in Section~\ref{lsm.sect}, we can construct a linear relationship between $\gamma_k$ and $\Gamma_k$,  
\be \label{linear1}
\Gamma_k = \beta_0 \gamma_k + \alpha_k, k = 1, \dots, p,
\ee
where $\alpha_k\sim \mathcal{N}(0,\sigma_\alpha^2)$, capturing  the direct effects of genetic variants on a health outcome. Combining Eqn. (\ref{ssdist1}) and (\ref{linear1}), various MR methods have been developed to estimate the causal effect $\beta_0$, e.g.,~RAPS~\cite{zhao2018statistical}, but they are all based on the InSIDE condition, namely, $\gamma_k$ and $\alpha_k$ are independent. If there is correlation between $\gamma_k$ and $\alpha_k$, causal effect $\beta_0$ is not identifiable in Eqn. (\ref{linear1}). Intuitively, this non-identifiability can be easily observed from a perspective of linear regression with correlation between the explanatory variable and residuals. 

\subsection{A motivating example}
\begin{figure}
	\centering
	\includegraphics[width=.8\linewidth]{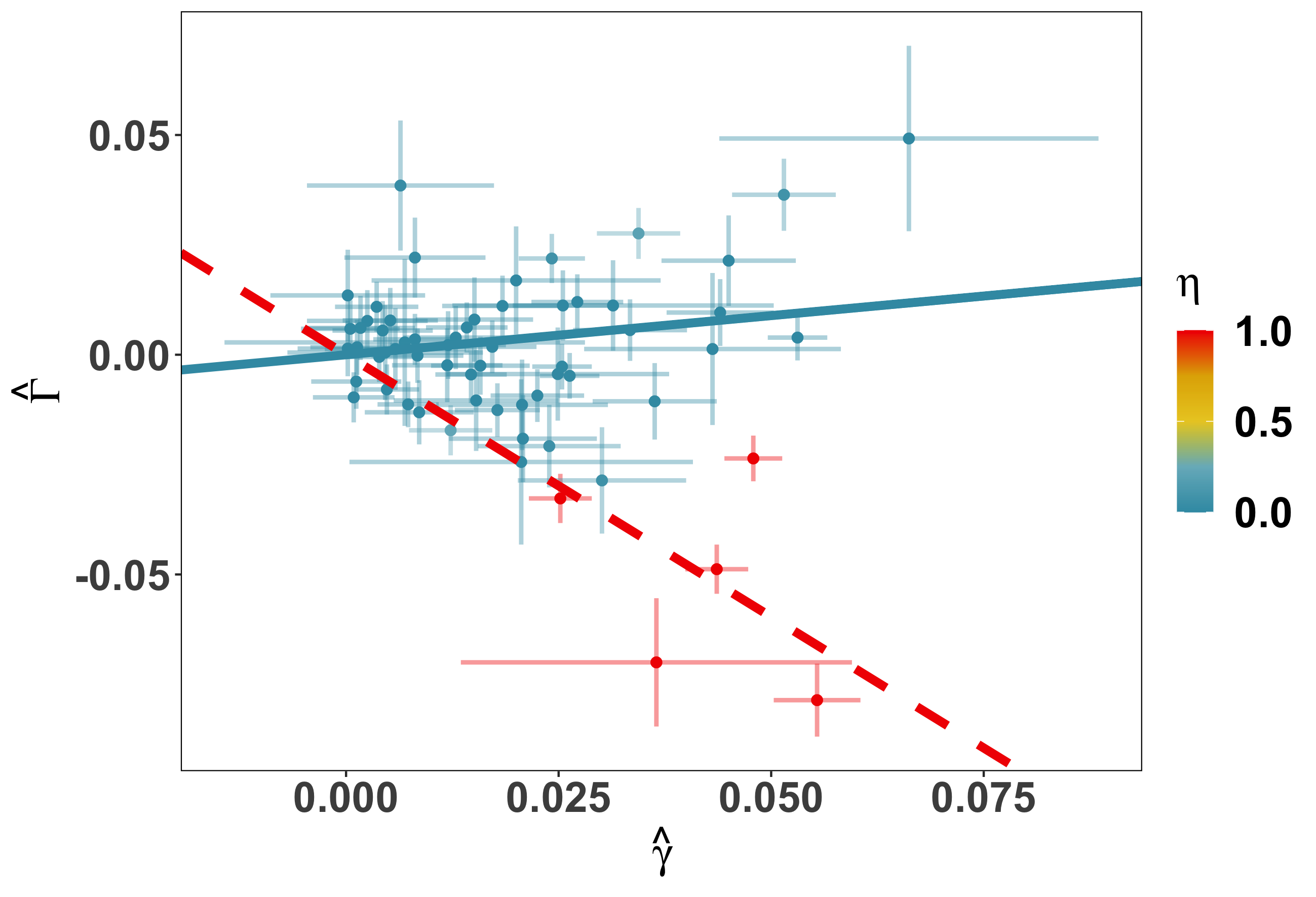}  %37
	\caption{Scatter plot of $\wh\Gamma_k$ against $\wh\gamma_k$ between CAD and HDL-c. The vertical and horizontal segments represent the standard errors of $\wh\Gamma_k$ and $\wh\gamma_k$, respectively, at each point. To ease the presentation, we orient the reference alleles so that $\wh\gamma_k$ is positive. For each data point, we use a spectrum of color bar to represent the mean value of latent indicators $\eta_k$. The blue line represents the causal effect $\hat\beta_0$ of HDL-c on CAD while the red line represents the nuisance parameter $\hat\beta_1$, see Eqn.~\ref{relat}.}\label{fig:motivate}
\end{figure}
Next we introduce a real data example that is used in the real data analysis. In this example, we are interested in estimating the causal effect of high-density lipoprotein cholesterol (HDL-c) on coronary artery disease  (CAD). We obtained summary statistics from three GWASs without overlapping samples, see Table S2 in the Supplementary document. 

Using HDL-c from Biobank Janpan~\cite{kanai2018genetic} as a screening dataset, we selected 62 SNPs that have $p$-value less than $1\times 10^{-6}$ and uncorrelated (through SNP clumping; \cite{purcell2007plink}). We then obtain the summary statistics $\{\wh \gamma_k, \wh \bfs_{\gamma_k}^2 \}$ and $\{\wh \Gamma_k, \wh \bfs_{\Gamma_k}^2 \}$, $k=1,\dots p$, for the SNP-HDL-c and and SNP-CAD, respectively. The scatter plot of $\wh\Gamma_k$ against $\wh\gamma_k$ for the 62 pairs of genetic effects with their standard errors is shown in Figure~\ref{fig:motivate}.  %The target is to estimate $\beta_0$ with potential horizontal pleiotropy. \
By assuming the systematic independent horizontal pleiotropy (IHP), a linear relationship~\ref{linear1} can be applied to estimate causal effect $\beta_0$. However, due to CHP, there exists heteroscedasticity in this linear relationship. With larger $\wh\gamma_k$, the variance of $\wh\Gamma_k$ becomes larger. The statistical method should overcome this heteroscedasticity to produce robust causal estimates.

\subsection{Methodological challenges and organization of the paper}
Similar to its related IV methods, how to correct for CHP or correlated direct effects remains the biggest challenge for MR methods using GWAS summary statistics. As discussed in Section~\ref{intro.sect1}, there exists non-identifiability issues when we use a bivariate normal distribution for $\gamma_k$ and $\alpha_k$. To tackle this issue, we propose a strategy using a mixture of linear regressions via orthogonal projection assuming the horizontal pleiotropy is sparse. In general, this strategy can be applied to other IV methods to mitigate the impact of correlated direct effects. 

Compared to IV methods and MR methods that use the individual-level data, most of classical MR methods for summary data cannot handle the correlations among instrumental variants.
%Due to the rate of genetic recombination, mutation rate and other factors, genetic variants across the genome present correlation patterns. 
Usually, two variants close in distance tend to have high correlations. Moreover, these correlations decay exponentially in genetic distance. Because of correlations among genetic variants, it is not valid to use i.i.d. normal distribution (\ref{ssdist1}) for GWAS summary statistics from a dense set of SNPs. Recently, we  developed MR-LDP~\cite{cheng2020mr} to infer causal effects using correlated weak instrumental variants accounting for both linkage disequilibrium and IHP. MR methods using correlated genetic variants will improve the statistical efficiency for point estimation. %The statistical efficiency will be improved if more valid genetic variants are used.  
To model summary statistics from correlated SNPs, the key is to use multivariate normal distributions to approximate the joint distributions of $\wh {\bfgamma}$ and $\wh {\bfGamma}$. 
The details of this approximated likelihood for GWAS summary statistics are given in the Supplementary document.

In this paper, we propose a unified and statistically efficient two-sample MR method to account for CHP using weak instrumental variants. To ease the presentation, we will consider the following two nested models using GWAS summary statistics:  \\%correlations in both horizontal pleiotropy and weak instruments,  MR-Corr$^2$
\indent MODEL 1 (MR-Corr). Accounting for CHP using independent instrumental variants.\\
\indent MODEL 2 (MR-Corr$^2$). Accounting  for CHP using correlated instrumental variants. 

%The consideration of these two models is based on the type of sparse indicators used.  In Section~\ref{mlr.sect}, we will introduce the strategy to correct for CHP using a mixture of linear regressions via orthogonal projection. As long as the instrumental variants are independent, we will assume an indicator $\eta_k$ for each variant $k$ with $\eta_k=1$ for $ \alpha_k \neq 0$ and $\eta_k=0$ for  $\wt \alpha_k  = 0 $. However, in the case of using correlated instrumental variants,  when there is a nonzero $\wt\alpha_k$ on the $k$-th variant, it is expected to have a nonzero horizontal pleiotropy $\wt\alpha_{k'}$ for  variant  $k'$ nearby the $k$-th variant because of the correlations between $G_k$ and $G_{k'}$. Fortunately, LD along the genome can be partitioned into independent blocks~\cite{berisa2016approximately} and thus variants from different blocks can be largely taken as independent. Because of partitioning the genome, a group-level indicator $\eta_l$, $l=1\dots, L$ can be used to identify blocks with zero orthogonal projected horizontal pleiotropy, where $L$ is the number of independent blocks. 

The rest of the paper is organized as follows. In Section~\ref{method}, we first review the linear structural models to estimate causal effects followed by introducing the mixture of linear regressions via orthogonal projection. We also introduce MR-Corr and MR-Corr$^2$ methods in this Section. %with some algorithmic details in this Section. 
In Section~\ref{connnect.sect}, we connect MR-Corr with existing methods. In Section~\ref{Validate}, %we first apply the proposed method for the analysis of a real validation using the height-height example, and then 
we perform simulation studies to benchmark the performance of  MR-Corr$^2$. In Section~\ref{real.data},
we first apply the proposed method for the analysis of a real validation using the height-height example, and then apply MR-Corr$^2$ to analyze GWAS summary statistics from two sets of traits. %ten traits.
%2, we show linear structural modeling  starting from no pleiotropy model to independent pleiotropy model. In Section 3, we introduce the proposed MR-Corr and MR-Corr$^2$ with an efficient MCMC algorithm. In Section 4, we show the related topics for summary statistics distribution, the choice of LD matrix, and connections with CAUSE. In Section 5, we show  simulation studies that MR-Corr$^2$ outperforms competing methods in terms of type-I error control and point estimates for making causal inference. In Section 6,, we first use two real exposure-outcome pairs to validate results from MR-Corr$^2$ in comparison with alternative methods, particularly showing our methods are more efficient using all variants in LD.% We will conclude the Introduction by briefly discussing the challenges in MR or IV methods and our solutions. %Therefore, we propose to use a group-level spike-slab 

\section{Methods}\label{method}
%To illustrate how MR methods work, we first start from linear structural models without/with (independent) horizontal pleiotropy, respectively, in Section~\ref{lsm.sect}.  Then, we show the MLR strategy to correct for correlated horizontal pleiotropy in Section~\ref{mlr.sect}. Using MLR strategy, we 
\subsection{Linear structural models and MR}\label{lsm.sect}
In this Section, we illustrate how MR methods work in the absence/presence of IHP using linear structural models~\cite{bowden2017framework, zhao2018statistical}. 
%Denote $\bfG\in\mathbb{R}^{n\times p}$ the genotype matrix for $p$ variants among $n$ independent individuals, $\bfx\in\mathbb{R}^{n\times 1}$ and $\bfy \in\mathbb{R}^{n\times 1}$ the corresponding exposure  and outcome variable, respectively, and $\bfU\in\mathbb{R}^{n\times q}$ the matrix for $q$ confounding factors among $n$ samples.  The SNP-exposure  and SNP-outcome true effects are denoted as $\bfgamma\in\mathbb{R}^{p\times 1}$ and $\bfGamma\in\mathbb{R}^{p\times 1}$, respectively, for all $p$ SNPs. 
\begin{figure}
	\centering
	\includegraphics[width=.8\linewidth]{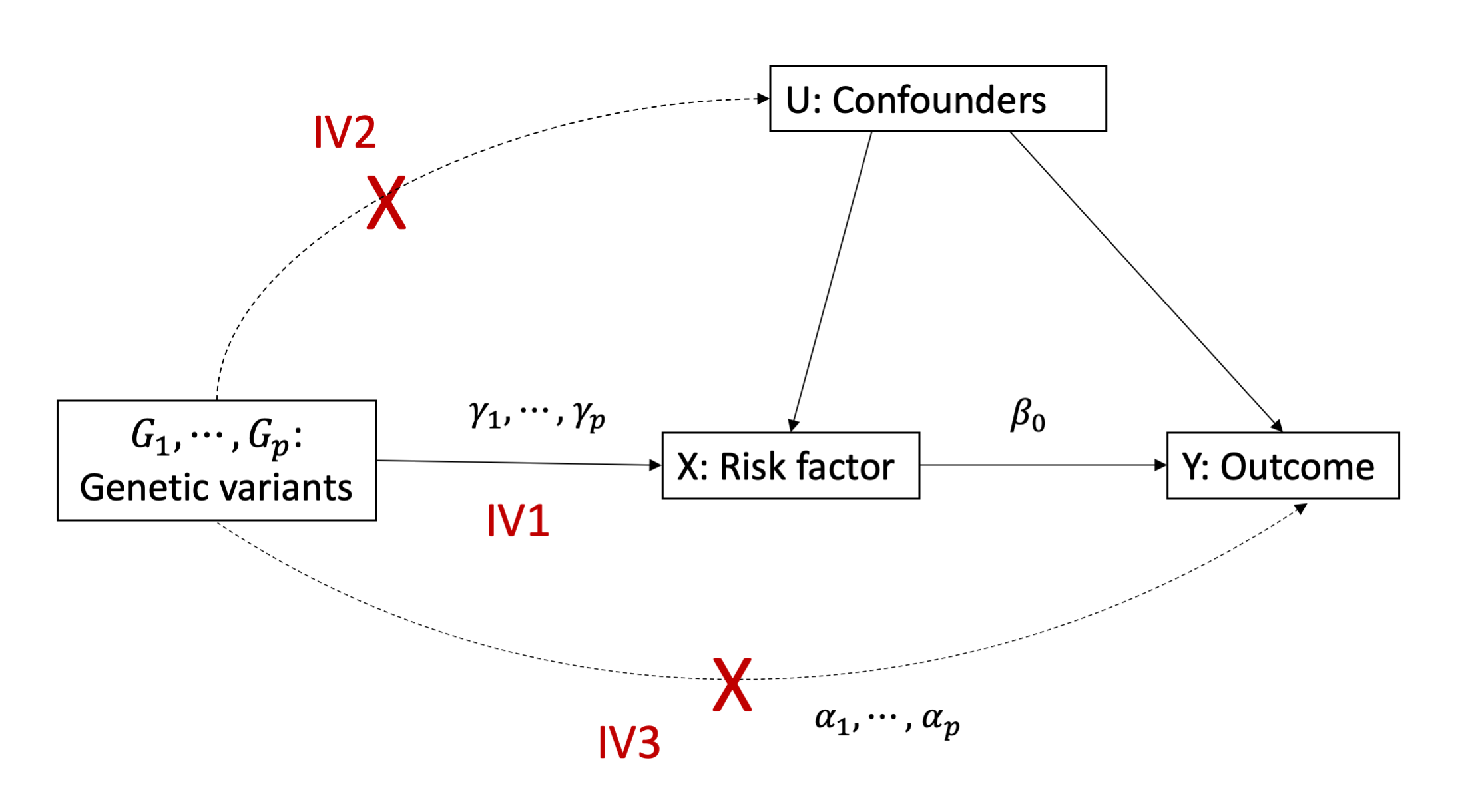}  %37
	\caption{Direct acyclic graph of MR and three IV assumptions. IV1: genetic variants ($G_k$s) are associated with the exposure (relevance); IV2: $G_k$s are not associated with unobserved confounders (independence); IV3: $G_k$s have no direct effects on the outcome (exclusion). To relax the exclusion assumption (IV3), most methods assume the InSide condition that $\gamma_k$ and $\alpha_k$ are independent.}\label{fig:dag}
\end{figure}

In MR analysis, we are interested in inferring causal relationship between an exposure $X$ and an outcome $Y$, with unobserved confounding factors $U$. 
The classical MR analysis obeys the following three assumptions in the IV methods, as shown in Figure~\ref{fig:dag}:
\begin{enumerate}
	{\setlength\itemindent{15pt} 	\item The {\sl relevance} assumption: The instrument $G$ is associated with the exposure $X$.}
	{\setlength\itemindent{15pt} 	\item The {\sl independence} assumption: $G$ is independent of confounders $U$.}
	{\setlength\itemindent{15pt} 	\item The {\sl exclusion} assumption: $G$ affects the outcome $Y$ only through the exposure $X$.}
\end{enumerate}
Assuming the validity of instrumental variants, we begin with the linear structural model,
\be\label{lsm1}
X =  \sum_{k = 1}^p  \gamma_k G_k   + \eta_x U + \epsilon_x, \quad Y = \beta_0 X + \eta_y U + \epsilon_y,
%\bfx =  \sum_{k = 1}^p  \bfg_k\gamma_k +  \bfU\bfeta_x + \bfepsilon_x, \quad \bfy =  \beta_0 \bfx +  \bfU \bfeta_y+ \bfepsilon_y,
\ee
where $\beta_0$ is the causal effect of interest, $\eta_x$ and $\eta_y$ are effects of confounding factors on exposure and outcome, respectively,  $\epsilon_x$ and $\epsilon_y$ are independent random noises,  $(\epsilon_x, \epsilon_y) \perp \!\!\! \perp (G_1,\dots,G_p,U)$, $\epsilon_x\perp \!\!\! \perp \epsilon_y$, and $U\perp \!\!\! \perp (G_1,\dots,G_p)$. In two-sample MR, we observe two independent samples with sample sizes $n_x$ and $n_y$ for $(X,G_1,\dots,G_p)$ and $(Y,G_1,\dots,G_p)$, respectively. Clearly, Eqn.  (\ref{lsm1}) satisfying the {\sl exclusion} assumption as $(G_1,\dots,G_p)\perp \!\!\! \perp Y | X $.  By replacing $X$ in the second equation with the first equation in Eqn. (\ref{lsm1}), we have 
\be
Y =  \sum_{k = 1}^p  \beta_0 \gamma_k G_k  +    ( \beta_0\eta_x + \eta_y) U + \beta_0 \epsilon_x + \epsilon_y. \notag
\ee
Therefore, the effect $\Gamma_k$ of variant $k$ on the health outcome can be written as 
\be\label{linear0}
\Gamma_k = \beta_0 \gamma_k, \forall k =1, \dots,p.
\ee 

In practice,  the {\sl exclusion} assumption can be easily violated because of ubiquitous horizontal pleiotropy or direct effects. Similar to Eqn. (\ref{lsm1}), the linear structural model incorporating horizontal pleiotropy  $\alpha_k$, on each variant $k$ can be written as  %We first look at following linear structural model with horizontal pleiotropy, $\alpha_k$, on each variant $k$,
\be\label{lsm2}
X = \sum_{k = 1}^p  \gamma_k G_k  +   \eta_x U+ \epsilon_x, \quad Y = \sum_{k = 1}^p  \alpha_kG_k+ \beta_0 X +  \eta_y U + \epsilon_y,
\ee
Again, by replacing $X$ in the second equation with the first equation in Eqn. (\ref{lsm2}), we have
\be
Y =  \sum_{k = 1}^p  (\beta_0 \gamma_k +\alpha_k)G_k+  ( \beta_0\eta_x + \eta_y)U+ \beta_0 \epsilon_x + \epsilon_y. \notag
\ee
Thus, the effect $\Gamma_k$ of variant $k$ on the health outcome can be written as Eqn. (\ref{linear1}).

In practice, performing causal inference using either Eqn. (\ref{lsm1}) or Eqn. (\ref{lsm2}) is impractical as it requires individual-level data, Fortunately, GWAS summary statistics are publicly accessible. 
Suppose that we obtain summary statistics for health risk factor and (disease) outcome as $\{\wh \gamma_k, \wh \bfs_{\gamma_k}^2 \}$ and $\{\wh \Gamma_k, \wh \bfs_{\Gamma_k}^2 \}, \forall k =1,\dots, p$, respectively, from two independent samples, where $p$ is the number of genetic variants. To ease the presentation, we first consider summary statistics for a set of independent genetic variants by SNP clumping. The corresponding MR-Corr model is introduced in Section~\ref{model.mrcorr}. Later, we extend our model to incorporate genetic variants within LD and introduce MR-Corr$^2$ in Section~\ref{model.mrcorr2}. For independent genetic variants, their distributions for summary statistics can be written as Eqn. (\ref{ssdist1}). %This distribution is guaranteed as long as the summary statistics $\{\wh \gamma_k, \wh \bfs_{\gamma_k}^2 \}$ and $\{\wh \Gamma_k, \wh \bfs_{\Gamma_k}^2 \}$ are calculated from independent samples. %From this point, various methods can be applied to estimate the causal effect $\beta_0$ using GWAS summary statistics based on Eqn. (\ref{linear0}), e.g., inverse variance weighting (IVW)~\cite{burgess2013mendelian}. 
To further account for horizontal pleiotropy, various MR methods have been developed to estimate the causal effect $\beta_0$ by assuming that $\bfalpha\perp \!\!\! \perp \bfgamma$ (the InSIDE condition). %e.g., RAPS~\cite{zhao2018statistical} and BMWR~\cite{zhao2020bayesian}. 
In genetic studies, the InSIDE condition may not hold due to heteroscedasticity in $\bfgamma$ and $\bfalpha$. In Section~\ref{connnect.sect}, we provide a perspective to treat CHP by connecting our methods with CAUSE. %Due to the complicated network among omics profiles, the phenomenon of CHP is pervasive for complex traits. %In general, this assumption is too strict to be satisfied in practice.

%\begin{remark}
{\it Remark 1}: From the linear structural model~(\ref{lsm1}), we know that when the classical IV assumptions (Figure~\ref{fig:dag}) hold, there exists a deterministic linear relationship between $\Gamma_k$ and $\gamma_k$, $\forall k = 1,\dots, p$, i.e., Eqn.~(\ref{linear0}). The resulting absolute correlation, denoted as $|\rho_g|$,  between $\bfGamma$ and $\bfgamma$ will be the exact 1, which is unrealistic as the absolute value of genetic correlation between any two distinct traits can never be 1. Note that $\rho_g$ is different from $\rho_{\alpha\gamma}$, which is the correlation between $\bfgamma$ and $\bfalpha$. On the other hand, the linear relationship with systematic IHP in Eqn.~(\ref{linear1}) is more flexible, resulting a correlation that $|\rho_g| < 1$. 
%\end{remark}

\subsection{Mixture of linear regressions via orthogonal projection}\label{mlr.sect}
To correct for CHP, we develop a strategy using a mixture of linear regressions (MLR) via orthogonal projection.  Our key idea is based on the following observation. For the ease of presentation, we assume that instrumental variants are independent here and will generalize this strategy to handle correlated instrumental variants in Section~\ref{model.mrcorr2}.  Suppose that the effects of genetic variant on exposure and horizontal pleiotropy follow a bivariate normal distribution for each variant $k$, %are correlated as bivariately normal distribution for each variant $k$,  
\be\label{bio}
\left(\begin{array}{c}
	\gamma_k\\
	\alpha_k
\end{array} \right)  \sim 
\N
\left( %\left(\begin{array}{c}
%	\gamma_k \\
%	\alpha_k
%\end{array} \right)
%\bigg|
%\left( \begin{array}{c}
%	0 \\
%	0
%\end{array} \right ),
\boldsymbol{0},
\left( \begin{array}{cc}
	\sigma_{\bfgamma}^2,&\rho_{\alpha\gamma}\ \sigma_{\bfgamma} \sigma_{\bfalpha_0}\\
	\rho_{\alpha\gamma}\ \sigma_{\bfgamma} \sigma_{\bfalpha_0 },&\sigma_{\bfalpha_0}^2 
\end{array} \right )
\right ),
\ee
where $\rho_{\alpha\gamma}$ is the correlation between $\gamma_k$ and $\alpha_k$. Clearly, from Eqn. (\ref{bio}), we have
\be\label{proj}
\gamma_k &\sim& \N(0, \sigma_{\bfgamma}^2), \notag\\
\alpha_k &=& \rho_{\alpha\gamma} \frac{\sigma_{\bfalpha 0}}{\sigma_{\bfgamma}} \gamma_k + \sqrt{1 - \rho_{\alpha\gamma}^2}  \sigma_{\bfalpha 0} Z_k \defby c \gamma_k + \wt \alpha_k, \notag
\ee
where $Z_k \perp \!\!\! \perp \gamma_k$ and $Z_k \sim \N(0, 1)$. In this way, $\alpha_k$ can be decomposed into two parts, one is in a linear relationship with $\gamma_k$ and the other is independent of $\gamma_k$, i.e., $\wt \alpha_k \perp \!\!\! \perp \gamma_k$ and $\wt \alpha_k \sim \N(0, \sigma_{\bfalpha}^2)$. We can further parameterize the causal relationship for genetic variant $k$ as follows
\be\label{relat}
\Gamma_k =  \left\{
\begin{aligned}
	&\beta_0 \gamma_k ,   &&\wt \alpha_k  = 0  \\
	&\beta_1 \gamma_k  +  \wt \alpha_k,  & &\wt \alpha_k \neq 0,
\end{aligned}
\right.
\ee
where $\beta_0$ is the causal effect of interest and $\beta_1=\beta_0+\rho_{\alpha\gamma} \frac{\sigma_{\bfalpha 0}}{\sigma_{\bfgamma}}$ is a nuisance parameter. Therefore, to remove the impact of CHP in Eqn. (\ref{bio}) is equivalent to identifying zero orthogonal projection of $\alpha_k$,  namely $\wt \alpha_k = 0$, in Eqn. (\ref{relat}). For this purpose, we may apply a spike-slab prior on $\wt \alpha_k$~\cite{ishwaran2005spike}. This observation motivates us to model $\Gamma_k$ as a mixture of two linear relationships with $\gamma_k$ depending on the value of $\wt \alpha_k$. Thus, using MLR strategy, the causal effect $\beta_0$ can be estimated consistently by removing the impact of CHP as long as correlation $\rho_{\alpha\gamma}$ is not too close to 1 or -1.   %propose a statistically joint model to perform hypothesis testing for causal inference about $\beta_0$ while accounting for the correlated pleiotropy by applying variable selection techniques on $\wt \alpha_k$.

\subsection{MR-Corr model}\label{model.mrcorr}
In practice, the InSIDE condition may not hold due to the pervasive heteroscedasticity among complex traits, %genetic correlations among complex traits~\cite{lee2013genetic,bulik2015atlas,pickrell2016detection}, 
resulting in correlations between horizontal pleiotropy, $\bfalpha$, and effects on exposure, $\bfgamma$. As shown in our simulations (Figure~\ref{fig:sim_rho02}), a small deviation from zero correlation may lead to severe inflation in type-I error, making MR methods not correcting for CHP less reliable. 

Here, we consider a Bayesian MR method via the MLR strategy to correct for CHP using independent instrumental variants. Assuming the independence among instrumental variants as well as independent samples for SNP-exposure and SNP-outcome, for each variant $k$, $\wh \gamma_k$ and $\wh \Gamma_k$ are i.i.d. distributed as Eqn.~(\ref{ssdist1}). By introducing a latent indicator $\eta_k$ for each variant $k$, we assign a spike-slab prior on  $\wt \alpha_k$~\cite{ishwaran2005spike, shi2019vimco},
\bse\label{spike-slab}
%\label{alpha.ss}
\wt\alpha_{k} \sim  \left\{
\begin{aligned}
	&\N(0, \sigma_{\bfalpha}^2) , &  &\eta_k  = 1\\
	&\delta_0(\alpha_{k}), &  &\eta_k = 0,
\end{aligned}
\right. 
\ese 
where $\N(0, \sigma_{\bfalpha}^2)$ denotes a normal distribution with mean 0 and variance $\sigma_{\bfalpha}^2$, $\delta_0$ denotes the Dirac delta function at zero,  $\eta_k=1$ means the $k$-th genetic variant present nonzero orthogonal projected  pleiotropic effect, and  $\eta_k=0$ means zero orthogonal projected  pleiotropic effect. Here, $\eta_k$ is a Bernoulli random variable with probability $\omega$ being 1, i.e., $\eta_k\sim\omega^{\eta_k}(1-\omega)^{1-\eta_k}$. %We further assume a Beta distribution for $\omega$, $\omega \sim \text{Beta}(a, b)$, and two inverse Gaussian priors for $\sigma_{\bfgamma}^2$ and $\sigma_{\bfalpha}^2$, i.e.,$ \sigma_{\bfgamma}^2 \sim \IG( a_{\bfgamma}, b_{\bfgamma})$ and  $\sigma_{\bfalpha}^2 \sim \IG( a_{\bfalpha}, b_{\bfalpha})$, respectively.

Using the reparameterization trick in Eqn. (\ref{relat}), the relationship between $\bfgamma$ and $\bfGamma$ can be constructed linearly as
\be\label{mlr}
\Gamma_{k} = \beta_0 (1- \eta_k) \gamma_{k} + \beta_1 \eta_k \gamma_{k} +  \eta_k\wt \alpha_{k},
\ee 
where $k=1,\dots,p$ and $ \bfgamma  \perp \!\!\! \perp \wt {\bfalpha}$. 
%For both $\beta_0$ and $\beta_1$, we assume a non-informative prior. Denote the hyperparameter space as $\bfTheta= \{ a_{\bfgamma}, b_{\bfgamma}, a_{\alpha}, b_{\bfalpha}, a, b\}$. 
Then, we propose the following hierarchical Bayesian model for MR-Corr with MLR strategy to account for CHP, 
\be\label{MR-Corr}
&  \wh \Gamma_k | \Gamma_k \stackrel{i.i.d.}{\sim} \mathcal{N}(\Gamma_k,\wh\bfs_{\Gamma_k}^2),\quad \wh \gamma_k | \gamma_k \stackrel{i.i.d.}{\sim} \mathcal{N}(\gamma_k,\wh\bfs_{\Gamma_k}^2), \notag\\
&  \Gamma_{k} = \beta_0 (1- \eta_k) \gamma_{k} + \beta_1 \eta_k \gamma_{k} +  \eta_k\wt \alpha_{k},\notag\\
& \gamma_k |\sigma_{\gamma}^2 \stackrel{i.i.d.}{\sim} \mathcal{N}(0,\sigma^2_{\gamma}), \quad \wt\alpha_k | \sigma_\alpha^2 \stackrel{i.i.d.}{\sim} \mathcal{N}(0,\sigma^2_{\alpha}),\quad \eta_k |\omega \stackrel{i.i.d.}{\sim}  \omega^{\eta_k}(1-\omega)^{1-\eta_k},\notag\\
& \sigma^2_\gamma \sim \mathcal{IG}(a_\gamma,b_\gamma), \quad  \sigma^2_\alpha \sim \mathcal{IG}(a_\alpha,b_\alpha), \quad \omega \sim Beta(a,b), 
\ee
where  Jeffreys non-informative priors  are used for both $\beta_0$ and $\beta_1$ and a spike-slab prior (\ref{spike-slab}) is reprameterized as $\wt\alpha_k\omega_k$. %~\cite{yang2018lpg}.  
%The hyperparameters for inverse Gamma distribution are ... {\color{red} discussion for the choice of hyperparamters.} 
The detailed Gibbs sampling algorithm with pseudo codes can be found in the Supplementary document.

\subsection{MR-Corr$^2$ model}\label{model.mrcorr2}
As MR-Corr~(\ref{MR-Corr}) assumes the SNPs are independent, it does not work for instrumental variants in LD. To further incorporate SNPs in LD for MR analysis, we develop MR-Corr$^2$ here. The key is to build a joint summary statistics  distribution for correlated SNPs. For this purpose, we use the approximated distribution for summary statistics~\cite{zhu2017bayesian,huang2018remi}, 
\be\label{ssdist}
\wh {\bfgamma}| \bfgamma, \wh \bfR, \wh \bfS_{\gamma} &\sim &\N(\wh \bfS_{\gamma} \wh \bfR \wh \bfS_{\gamma}^{-1} \bfgamma, \wh \bfS_{\gamma} \wh \bfR \wh \bfS_{\gamma}) ,\notag\\
\wh {\bfGamma}| \bfGamma, \wh \bfR, \wh \bfS_{\Gamma} &\sim &\N(\wh \bfS_{\Gamma} \wh \bfR \wh \bfS_{\Gamma}^{-1} \bfGamma,\wh \bfS_{\Gamma} \wh \bfR \wh \bfS_{\Gamma}) ,
\ee
where $\wh {\bfS_{\gamma} }= \text{diag}([\wh \bfs_{\gamma_1}, \cdots, \wh \bfs_{\gamma_p}] )$  and $\wh{ \bfS_{\Gamma}}= \text{diag}([\wh \bfs_{\Gamma_1}, \cdots, \wh \bfs_{\Gamma_p}] )$  are both diagonal matrices, $\wh {\bfgamma }=  [\wh \gamma_1,\dots,\wh \gamma_p]^T$, $\wh{ \bfGamma }=  [\wh \Gamma_1,\dots,\wh \Gamma_p]^T$, $\bfgamma =  [ \gamma_1,\dots, \gamma_p]^T$, $\bfGamma =  [ \Gamma_1,\dots, \Gamma_p]^T$. Given that sample size $n$ is large enough and the trait is highly polygenic, i.e., the squared correlation coefficient between the trait and each genetic variant is close to zero, Zhu and Stephens~\cite{zhu2017bayesian} showed that difference between the likelihoods based on the individual-level data %(\ref{ind_level}) 
and summary statistics %(\ref{sum0}) 
is some constant that does not depend on $\bfgamma$ under some regularity conditions (See Table 1 in~\cite{zhu2017bayesian} and Table S4  in~\cite{yang2020comm} for numerical justifications). Details for this approximated distribution can be found in Section S1.1. %~\ref{ss.sect}. 

As GWAS summary statistics does not contain any information for the correlation $\wh \bfR$,  we use an additional  independent sample as reference data to estimate this correlation $\wh \bfR^{\mathrm{ref}}$. In \cite{huang2018remi}, we theoretically analyzed the impact of using reference sample to estimate this correlation for $l_1$ penalization and empirical results in~\cite{zhu2017bayesian, huang2018remi, cheng2020mr} show that the distribution for summary statistics approximates the individual-level one well. The details for the choice of LD can be found in Section Section S1.2.%~\ref{LDChoice}. 

Next, we show how to use MLR strategy for correlated SNPs. Clearly, when there is a single variant $k$ with nonzero direct effect $\alpha_k$, the direct effect for a nearby variant $k'$ would be nonzero as well. This is because variants $k$ and $k'$ from the same LD are highly correlated. Moreover, LD across the genome can be partitioned into independent blocks. SNPs within the same block are correlated but SNPs from different blocks are independent. Thus, the orthogonal projection $\wt\alpha_k$ should be in a group manner. Here, we introduce a group-level latent status $\eta_l$, indicating whether genetic variants within the $l$-th block present nonzero horizontal pleiotropy and assign a spike-slab prior on $\wt\alpha_{lk}$~\cite{ishwaran2005spike, shi2019vimco},
%\be\label{relatg}
%\wt\alpha_{lk} \sim  \left\{
%\begin{aligned}
%	&\N(0, \sigma_{\bfalpha}^2) , &  &\eta_l  = 0\\
%	&\delta_0(\alpha_{lk}), &  &\eta_l = 1,
%\end{aligned}
%\right. 
%\ee
\be\label{relatg}
\wt\alpha_{lk} \sim  \left\{
\begin{aligned}
	&\N(0, \sigma_{\bfalpha}^2) , &  &\eta_l  = 1\\
	&\delta_0(\alpha_{lk}), &  &\eta_l = 0,
\end{aligned}
\right. 
\ee

where $\N(0, \sigma_{\bfalpha}^2)$ denotes a normal distribution with mean 0 and variance $\sigma_{\bfalpha}^2$, $\delta_0$ denotes the Dirac delta function at zero,  $\eta_l=1$ means variants within the $l$-th block present nonzero orthogonal projected  pleiotropic effects or  $\eta_l=0$ means orthogonal projected  pleiotropic effects from variants within block $l$ are all zero. Here, $\eta_l$ is a Bernoulli random variable with probability $\omega$ being 1, $\eta_l\sim\omega^{\eta_l}(1-\omega)^{1-\eta_l}$. %We further assume a Beta distribution for $\omega$, $\omega \sim \text{Beta}(a, b)$, and two inverse Gaussian priors for $\sigma_{\bfgamma}^2$ and $\sigma_{\bfalpha}^2$, i.e.,$ \sigma_{\bfgamma}^2 \sim \IG( a_{\bfgamma}, b_{\bfgamma})$ and  $\sigma_{\bfalpha}^2 \sim \IG( a_{\bfalpha}, b_{\bfalpha})$, respectively.

Using the reparameterization trick in Eqn. (\ref{relatg}), the relationship between $\bfgamma$ and $\bfGamma$ can be constructed linearly as:
\be
%\Gamma_{lk} = \beta_0 \eta_l \gamma_{lk} + \beta_1 (1 - \eta_l) \gamma_{lk} +(1 - \eta_l)\wt \alpha_{lk},
\Gamma_{lk} = \beta_0(1 - \eta_l)  \eta_l \gamma_{lk} + \beta_1\eta_l \gamma_{lk} + \eta_l\wt \alpha_{lk},
\ee 
where $l=1,\dots,L$, $k=1,\dots,p_l$ and $\gamma_{lk}\perp \!\!\! \perp \wt \alpha_{lk}$. Then we propose the following hierarchical Bayesian model for MR-Corr$^2$,
\be\label{MR-Corr2}
&  \wh {\bfgamma}| \bfgamma, \wh \bfR, \wh \bfS_{\gamma} \sim \N(\wh \bfS_{\gamma} \wh \bfR \wh \bfS_{\gamma}^{-1} \bfgamma, \wh \bfS_{\gamma} \wh \bfR \wh \bfS_{\gamma}) ,\notag\\
& \wh{ \bfGamma}| \bfGamma, \wh \bfR, \wh \bfS_{\Gamma} \sim \N(\wh \bfS_{\Gamma} \wh \bfR \wh \bfS_{\Gamma}^{-1} \bfGamma,\wh \bfS_{\Gamma} \wh \bfR \wh \bfS_{\Gamma}) ,\notag\\
& \Gamma_{lk} = \beta_0(1 - \eta_l)  \eta_l \gamma_{lk} + \beta_1\eta_l \gamma_{lk} + \eta_l\wt \alpha_{lk}, \notag\\ % \Gamma_{lk} = \beta_0 \eta_l \gamma_{lk} + \beta_1 (1 - \eta_l) \gamma_{lk} +(1 - \eta_l)\wt \alpha_{lk},\notag\\
&{\color{black} \gamma_{lk} |\sigma_{\gamma}^2 \stackrel{i.i.d.}{\sim} \mathcal{N}(0,\sigma_{\gamma}^2), \quad \wt\alpha_{lk} | \sigma_\alpha^2 \stackrel{i.i.d.}{\sim} \mathcal{N}(0,\sigma_{\alpha}^2),
	\quad \eta_l |\omega \stackrel{i.i.d.}{\sim}  \omega^{\eta_l}(1-\omega)^{1-\eta_l}, }\notag \\ 
&  \sigma^2_\gamma \sim \mathcal{IG}(a_\gamma,b_\gamma), \quad  \sigma^2_\alpha \sim \mathcal{IG}(a_\alpha,b_\alpha), \quad \omega \sim Beta(a,b).
\ee
The corresponding parallel Gibbs sampling algorithm with pseudo codes can be found in the Supplementary document.

\section{Connections with Existing Methods}\label{connnect.sect}

\begin{figure} [ht!]
	\begin{subfigure}{.5\textwidth}
		\centering
		% include first image
		\includegraphics[width=.99\linewidth]{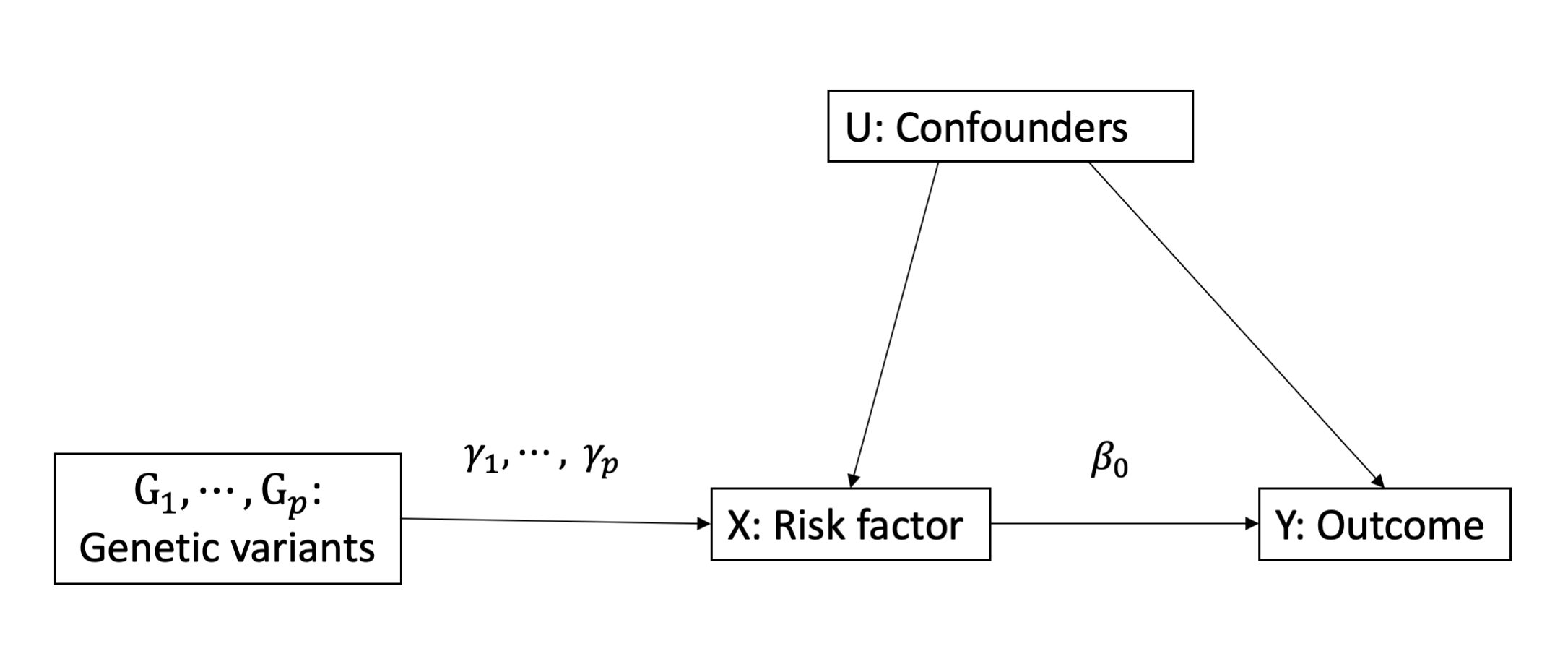}  %34
		\caption{Mechanism 1}
		\label{fig:mech1}
	\end{subfigure}	
	\begin{subfigure}{.5\textwidth}
		\centering
		% include first image
		\includegraphics[width=.99\linewidth]{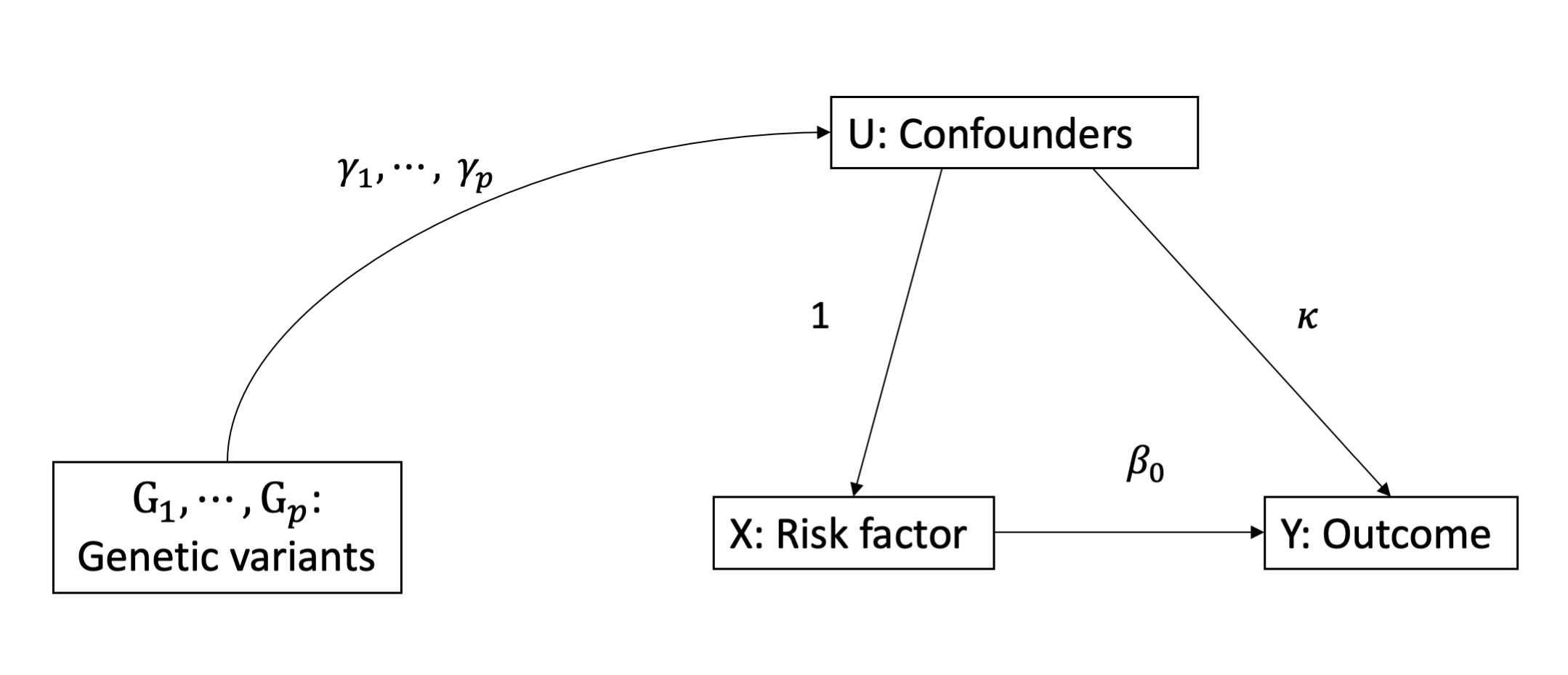}  %10
		\caption{Mechanism 2}
		\label{fig:mech2}
	\end{subfigure}	
	\caption{The causal mechanisms between exposure $X$ and outcome $Y$. Mechanism 1 is the classical one while mechanism 2 refers to correlated pleiotropy.}\label{fig:dag2}
\end{figure}

Recently, \cite{morrison2020mendelian} proposed a CAUSE method to address the problem that genetic variants affect both the exposure and outcome through a heritable shared factor and thus there are two mechanisms that SNPs affect an outcome as shown in Figure~\ref{fig:dag2}. Here, to ease the illustration, we consider only correlated pleiotropy for independent SNPs as shown in  Figure~\ref{fig:mech2}. 
By assuming the effect of confounding factor $U$ on exposure $X$ is scaled to 1 and considering the effect of $U$ on $Y$ is $\kappa$, the relationship between $\gamma_k$ and $\Gamma_k$ without considering horizontal pleiotropy can be written as
\be\label{cause.mech}
\Gamma_k = \beta_0 \gamma_k + \eta_k \kappa \gamma_k,
\ee
where %$\alpha_k$ is an uncorrelated horizontal pleiotropy and 
$\eta_k$ is an indicator of whether $G_k$ affects $U$ that has different meaning in MR-Corr. %When $\omega_k=1$, we have $\Gamma_k = \beta_0 \gamma_k +  \kappa \gamma_k + \theta_k$ for Eqn.~\ref{cause.mech}. 
Using the fact that  $\beta_1= \beta_0+\kappa$, simple algebra shows that  when $\wt \alpha_k \rightarrow 0$,  Eqn.~(\ref{mlr}) reduces to  Eqn.~(\ref{cause.mech}).  Thus, the CAUSE model with only correlated pleiotropy can be taken as an extreme case of MR-Corr. By using the mechanism of CHP, %mechanism of correlated pleiotropy in CAUSE (Figure X), 
the term $\kappa\gamma_k$ in Eqn.~(\ref{cause.mech}) can be taken as horizontal pleiotropy but is fully correlated with $\gamma_k$ with $|\rho_{\alpha\gamma}|=1$. This only happens when there is a single confounding factor $U$ that affects both the exposure and outcome. In this case, $\wt \alpha_k$ = 0 exactly as long as this single confounder has a scaled effect $\kappa$ on $Y$. However, when there are multiple confounding factors ($U_1,\dots,U_q)$ and these confounding factors affect the outcome through some mediators as well as present direct effects, the relationship between these confound factors and outcome $Y$ would therefore show deviations from the deterministic linear relationship with random errors. These random errors could represent the independent direct effects between multiple confounders and outcome. In this case, $\wt \alpha_k$ would not be 0 and Eqn.~(\ref{mlr}) holds. %In this sense, CAUSE model deals with only an extreme case for correlated pleiotropy. 

Moreover, the identifiability of CAUSE model is up to the identifiability of indicator $\eta_k$. Moreover, the consistency of estimating causal effects $\beta_0$ depends on how well one can estimate $\eta_k$. Compared to identify  which  mechanism is applied to a genetic variant as in CAUSE, identifying whether $\wt \alpha_k$ is 0 is much easier and also identifiable. As our methods avoid the case that $|\rho_{\alpha\gamma}|$ lies on the boundary, there will be no identifiablity problems.   %Similarly, MR-Corr and MR-Corr$^2$ only work for the case that $|\rho|\ne 1$ and in this case, the idenfiability and consistency can be guaranteed as long as   %this model would exhibit identifiability issue if $\omega_k$ is not identifiable.    %As $\eta_k$ in MR-Corr is the indicator for $\wt \alpha_k$ being 0,  %Moreover, $\rho \frac{\sigma_{\alpha 0}}{\sigma_{\gamma}}$ in MLR strategy can be interpreted as scaled effects of confounding factors $U$ on the outcome, $\kappa$ in CAUSE model. 

%On the other hand, according to MLR (Eqn.~\ref{mlr}), when $\eta_k=1$, we have 
%\be\label{mlr.eta1}
%\Gamma_k = \beta_1\eta_k\gamma_k+\wt \alpha_k,
%\ee
%where $\beta_1=\beta_0+\rho \frac{\sigma_{\bfalpha 0}}{\sigma_{\bfgamma}}$ and $\rho \frac{\sigma_{\bfalpha 0}}{\sigma_{\bfgamma}}$ collapses all information for $\kappa$ in Eqn.~\ref{cause.mech}. However, the identifiability is only for $\beta_1$ but not $\beta_0$ using   Eqn.~\ref{mlr.eta1} that makes CAUSE model~\ref{cause.mech} problematic. Using MLR strategy (Eqn.~\ref{mlr}), one would also reduce the impact of correlated pleiotropy in the sense of CAUSE, which was demonstrated  in the simulation studies.
%{\color{red} More discussions.}

Because of the existence of CHP, in many cases, one would observe that many SNPs do not obey either Eqn.~(\ref{linear0}) or Eqn.~(\ref{linear1}). RAPS and BWMR tackle this issue by treating these SNPs as idiosyncratic outliers. Despite the fact that estimating causal effect using either Eqn.~(\ref{linear0}) or Eqn.~(\ref{linear1}) is similar to performing a linear regression, treating SNPs that do not obey these relationships as idiosyncratic outliers is not justified. This is because we work on the true effect sizes $\bfgamma$ and $\bfGamma$ without measurement errors. There should exist mechanisms that cause this phenomenon and CHP is one of them. As one can observed in Figure~\ref{fig:motivate}, when the value of $\wh\gamma_k$ is large, the variation of $\wh\Gamma_k$ is large, causing the variance of ``residuals" to increase with the 'predictor variable' $\wh\gamma_k$. In the literature on linear regression, variance-stabilizing transformation can be applied, e.g. Box-Cox transformation~\cite{sakia1992box}. However, this technique is not applicable in MR as we work on unobserved estimated effects   $\wh\bfgamma$ and $\wh\bfGamma$ instead of the true effects of them. Fortunately, only SNPs with CHP would present this problem as  idiosyncratic outliers and MLR strategy can effectively mitigate the impact from these SNPs.

\section{Simulation Studies}\label{Validate}
%\subsection{Simulations}
\subsection{Simulation settings}\label{sim.setting}
In this section, we performed simulation studies to evaluate the performance of MR-Corr$^2$. %For summary statistics from both independent and correlated SNPs, 
In this simulation, we considered two scenarios: 1) The generative model for individual-level data generated as Eqn.~\ref{sim.gen} with correlated $\bfalpha$ and $\bfgamma$; 2) The correlated pleiotropy model in CAUSE setting as shown in Figure~\ref{fig:dag2}. 

In Scenario 1, we used the following structural model to generate individual-level data%the individual-level data from two independent samples can be generated as follows
\be\label{sim.gen}
\bfx = \bfG_1\bfgamma +\bfU_x  \bfeta_x  + \bfe_1,\quad \bfy = \beta_0 \bfx + \bfG_2 \bfalpha +  \bfU_y\bfeta_y+ \bfe_2,
\ee
where $\bfG_1 \in \mathbb{R}^{n_1\times p}$ and $\bfG_2 \in \mathbb{R}^{n_2\times p}$ are genotype  matrices for two samples, {\color{black} $\bfU_x \in \mathbb{R}^{n_1\times r }$ and $\bfU_y \in  \mathbb{R}^{n_2\times r}$} are matrices for confounding variables,  $n_1$ and $n_2$ are the corresponding  sample sizes that are set at 20,000, $p$ is the number of genetic variants, $\bfx\in \mathbb{R}^{n_1\times 1}$ is the exposure vector, $\bfy\in \mathbb{R}^{n_2\times 1}$ is the outcome vector, and $\bfe_1$ and $\bfe_2$ are independent random errors. %were obtained from $\N(\bf0, \sigma_{\bfe_1}^2\bfI_{n_1})$ and $\N(\bf0, \sigma_{\bfe_2}^2\bfI_{n_2})$, respectively. 
Note that Eqn.~(\ref{sim.gen}) is the sample version of the linear structural model~(\ref{lsm2}). 

In Scenario 2, we first generated $q$ from $Beta(1,10)$ as in~\cite{morrison2020mendelian}. Then we divided all $p$ genetic variants into two groups corresponding to two mechanisms in Figure \ref{fig:dag2}(a) and (b), respectively, with proportion {\color{black} $1-q$ and $q$}. For genetic variants with {\color{black}proportion  $1-q$} in mechanism 1 (Figure~\ref{fig:dag2}(a)), the exposure vector was generated as 
\bse
\bfx^{1} = \bfG_1^1\bfgamma^1 +\bfU_x^1  \bfeta_x^1 + \bfe_1^1,
\ese
where subscript indicates which mechanism is active, $\bfG_1^1$ is the genotype matrix in SNP-exposure sample for the genetic variants from mechanism 1 and $\bfgamma^1$ is the corresponding coefficient vector. For the remaining SNPs, the exposure  vector was generated as 
\bse
\bfU_x^2 = \bfG_{1}^2\bfgamma^2 + \bfe_{\bfU},\quad
\bfx^2 = \bfU_x^2 + \bfe_{1}^2, 
\ese
where $\bfG_1^2$ is the genotype matrix in SNP-exposure sample for the genetic variants from mechanism 2 and $\bfgamma^2$ is the corresponding coefficient vector. By combining these two mechanisms, the outcome vector can be generated as 
\bse
\bfx = \bfx^{1} + \bfx^2,\quad \bfy = \beta_0 \bfx + \bfG_2 \bfalpha +  \bfU_y\bfeta_y+ \bfe_2.
\ese

For both scenarios, we first generated genotype matrices from multivariate normal distribution $\N(\bf0, \bfSigma(\rho))$,  where $\bfSigma(\rho)$ is a block autoregressive (AR) with $\rho = 0.4$, or $0.8$ representing moderate or strong LD, respectively. The genotype matrices were then categorized into
dosage values $\{0, 1, 2\}$ according to the Hardy-Weinberg equilibrium using the minor allele frequencies drawn  from a uniform distribution $\mathbb{U}(0.05,0.5)$.  Moreover, we assumed that $\gamma_k$ and $\alpha_k$ from a bivariate normal distribution $\N(\bf0, \bfSigma(\rho_{\alpha\gamma}))$, where $\rho_{\alpha\gamma}$ is the correlation between $\alpha_k$ and $\gamma_k$. We varied $\rho_{\alpha\gamma}\in\{0, 0.2. 0.4\}$.  In addition, we considered $\alpha$ to be sparse, i.e.,  only a fraction of $\alpha_k$ was from this bivariate normal distribution and others are zero. The sparsity was fixed at 10\%. %that  only a fraction of $\alpha_k$ was from this bivariate normal distribution others remained 0, the sparsity was fixed at 0.1.

For confounding variables, we sampled each column of $\bfU_x$ and $\bfU_y$ from a standard normal distribution with fixed $r=50$. For their corresponding coefficients $\bfeta_x \in  \mathbb{R}^{r\times 1}$ and $\bfeta_y \in \mathbb{R}^{r \times 1}$, each row of $(\bfeta_x, \bfeta_y)$ was generated from a multivariate normal distribution $\N(\bf0, \bfSigma_{\eta})$, where $\bfSigma_{\eta}$ is a two-by-two matrix with diagonal elements set as 1 and  off-diagonal elements set as 0.8.
% which has $q$  and $1-q$ proportion  variants corresponded two mechanisms in Figure \ref{fig:dag2} (a) and (b), respectively. 
%In the first scenario, we first generated the individual 

We then conducted single-variant analysis to obtain the summary statistics for SNP-exposure and SNP-outcome, $\{\wh \gamma_k, \wh s_{\bfgamma k}^2 \}_{k = 1, \cdots, p}$ and $\{\wh \Gamma_k, \wh s_{\bfGamma k}^2\}_{k = 1, \cdots, p}$, respectively. In simulations, we controlled the signal magnitude for both $\bfgamma$ and $\bfalpha$ using their corresponding heritability, $h_{\bfgamma}^2 =\frac{\var (\beta_0\bfG_1\bfgamma)}{\var(\bfy)}$ and $h_{\bfalpha} ^2= \frac{\var(\bfG_2\bfalpha)}{\var(\bfy)}$, respectively. Thus, we could control $h_{\bfalpha} ^2$ and $h_{\bfgamma}^2 $ at any value by controlling confounding variables, and the error terms, $\sigma_{\bfe_1}^2$ and $\sigma_{\bfe_2}^2$.  In all settings, we fixed $h_{\bfgamma}^2 = 0.1$ and varied $h_{\bfalpha} ^2 \in \{ 0.05, 0.1\}$. 

\subsection{Simulation results}

We benchmark the performance of MR-Corr$^2$ using correlated SNPs, i.e., $\rho=0.4$ and $0.8$, in comparison with alternative methods, including MR-LDP, CAUSE, GSMR, MRMix and RAPS. In this simulation, we set the number of blocks $L$ to be 100 or 200 and the number of SNPs within a block to be 10. Thus we have $p$= 1,000 or 2,000 in total. As MR-Corr$^2$ and MR-LDP need an additional reference data to estimate LD, we generated an independent reference panel $G_3$ with sample size 500. Next we applied all six methods on the simulated  summary statistics as shown in Section~\ref{sim.setting}. 
Since MR-Corr, CAUSE, GSMR, MRMix and RAPS cannot account for strong correlated SNPs, we conducted SNP clumping to obtain independent SNPs for making fair comparisons on point estimates. 

\begin{figure} [ht!]
	\begin{subfigure}{.5\textwidth}
		\centering
		% include first image
		%\includegraphics[width=.99\linewidth]{Figure/MR2.png}  %34
		\includegraphics[width=.99\linewidth]{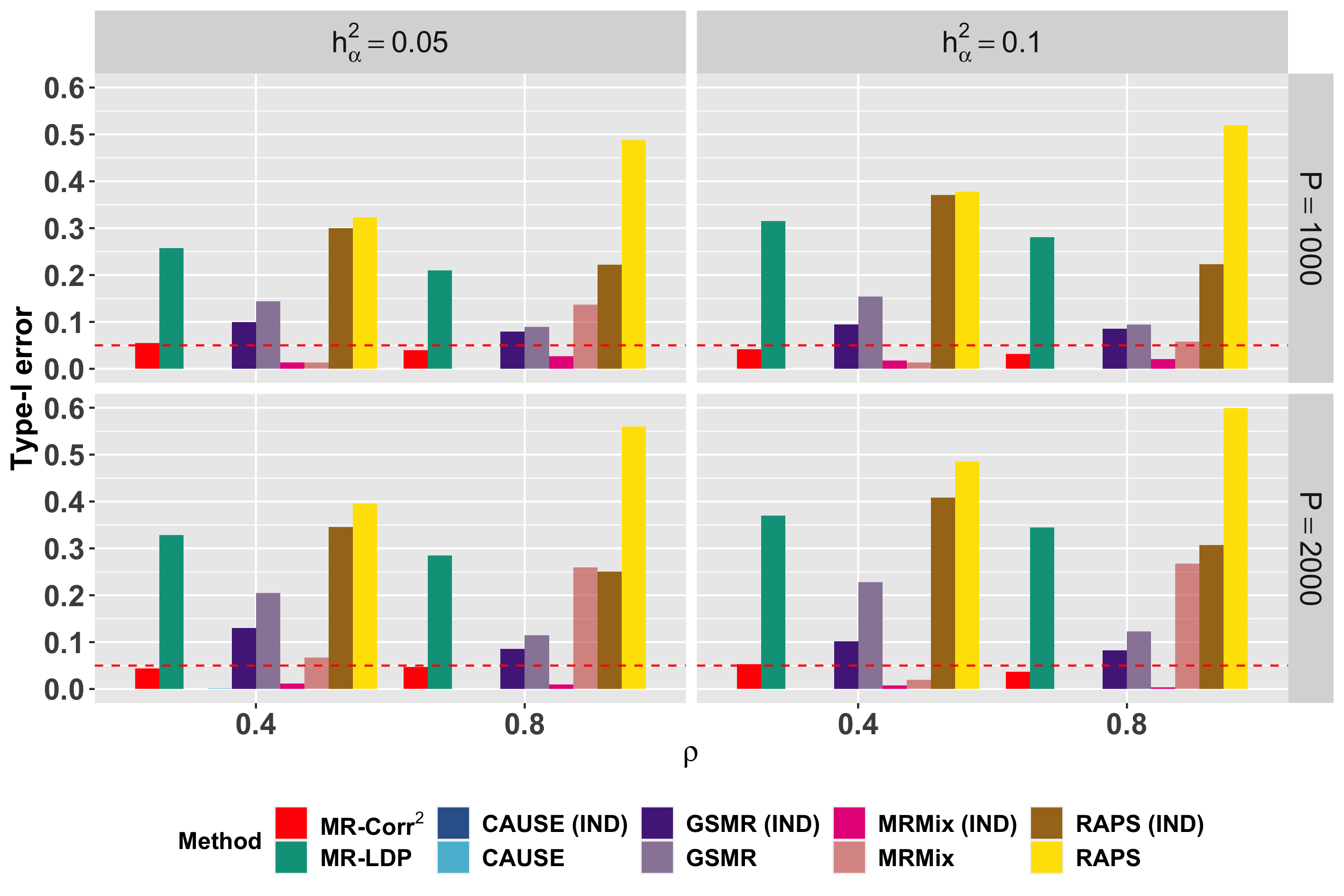}  %
		\caption{Type-I error}
		\label{fig:t1e_ld_g}
	\end{subfigure}	
	\begin{subfigure}{.5\textwidth}
		\centering
		% include first image
		\includegraphics[width=.99\linewidth]{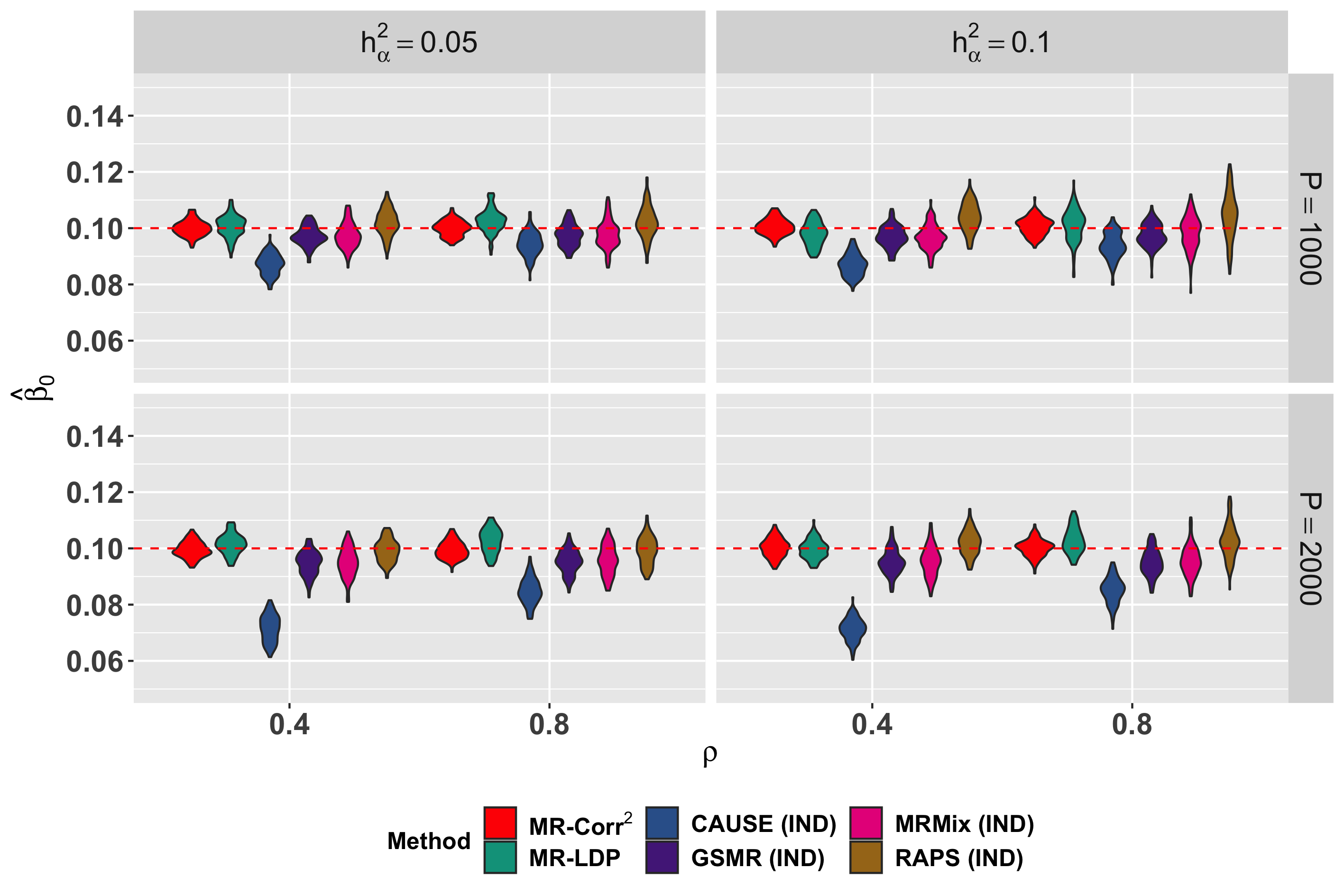}  %
		\caption{Point estimate}
		\label{fig:pe_ld_g}
	\end{subfigure}	
	\begin{subfigure}{.5\textwidth}
		\centering
		% include first image
		%\includegraphics[width=.99\linewidth]{Figure/MR2.png}  %34
		\includegraphics[width=.99\linewidth]{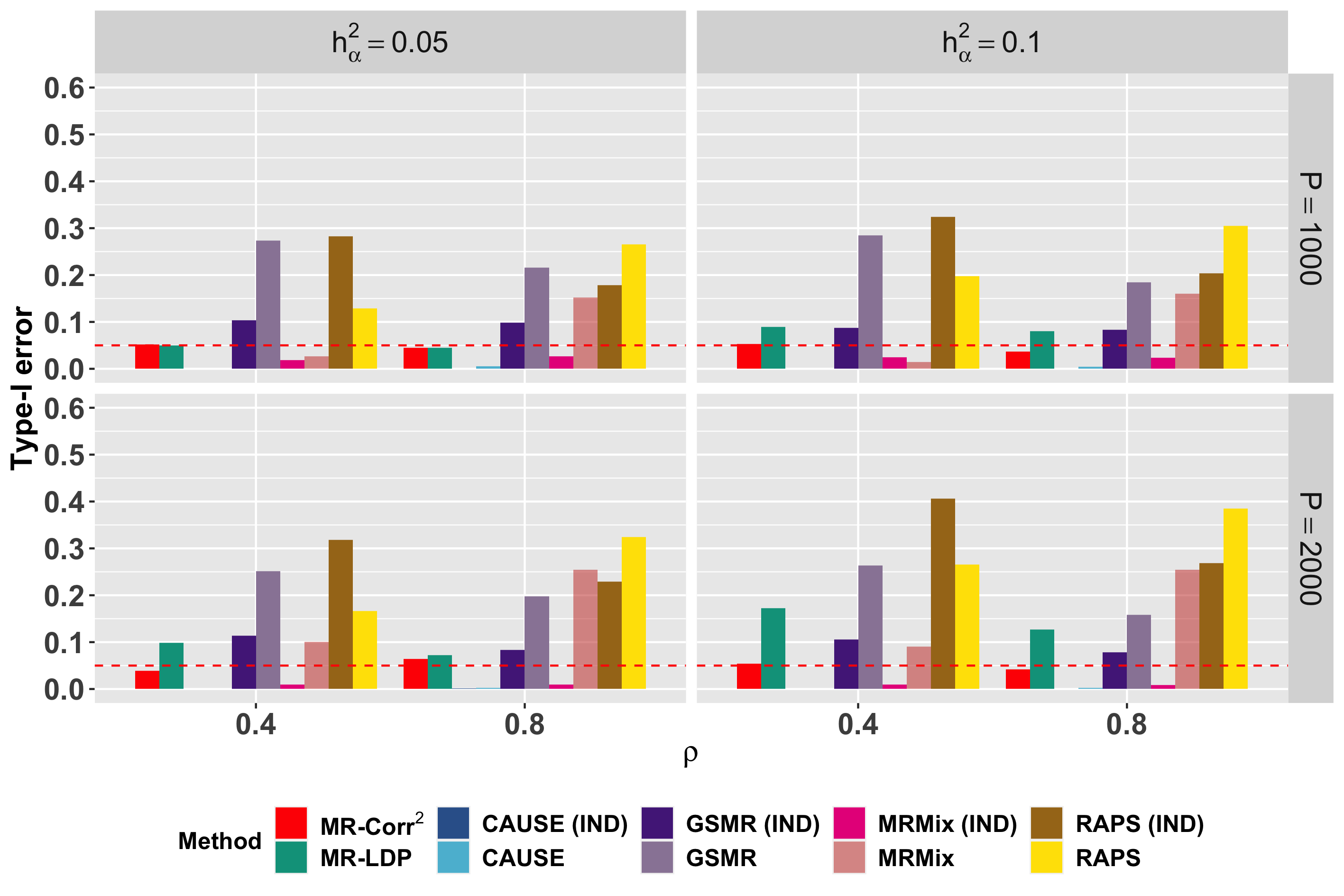}  %34
		\caption{Type-I error}
		\label{fig:t1e_ld_c}
	\end{subfigure}	
	\begin{subfigure}{.5\textwidth}
		\centering
		% include first image
		\includegraphics[width=.99\linewidth]{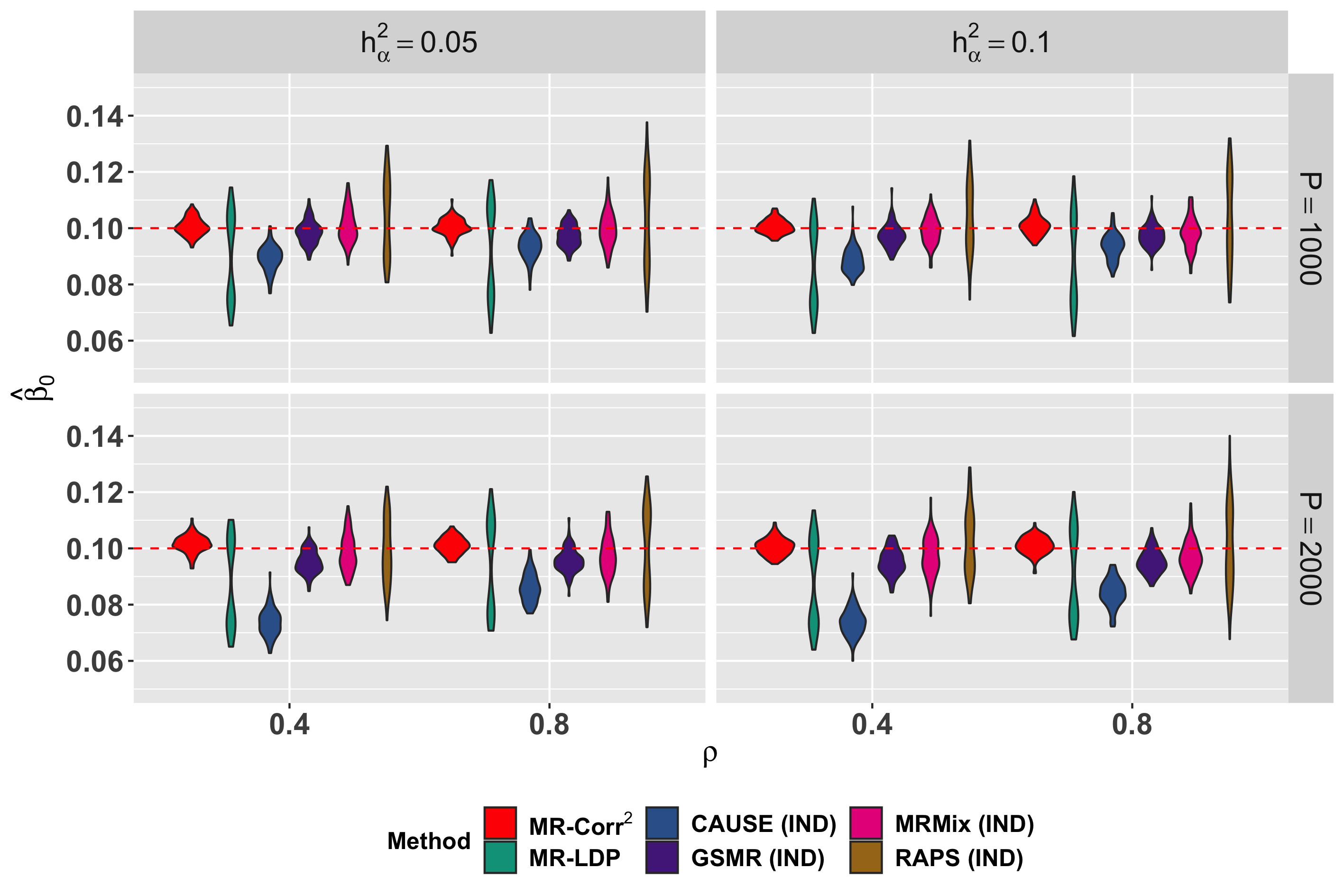}  %10
		\caption{Point estimate}
		\label{fig:pe_ld_c}
	\end{subfigure}	
	\caption{Comparisons of type-I error and point estimate using correlated SNPs with $\rho_{\alpha \gamma} = 0.2$. The number of replicates for type-I error is 1,000 and that for point estimates is 100. The top and bottom panels are corresponding to Scenario 1 and Scenario 2, respective. Note that we evaluate type-I error for CAUSE, GSMR, MRMix and RAPS using either all SNPs or independent SNPs. }\label{fig:sim_rho02}
\end{figure}

In different scenarios, we evaluated type-I error under the null that $\beta_0=0$ while we set $\beta_0=0.1$ to evaluate point estimates under the alternative. We ran 1,000 and 100 replicates to assess type-I error and point estimates, respectively. Figure~\ref{fig:sim_rho02} shows type-I error and point estimates of all methods when $\rho_{\alpha \gamma} = 0.2$. As we can see, MR-Corr$^2$ can effectively control type-I error in the presence of LD with CHP in both scenarios. While the type-I error of MR-LDP is a little inflated in the second scenario, it is severely inflated in the case of Scenario 1. Using all correlated SNPs, type-I errors for GSMR, MRMix and RAPS are severely inflated. After performing SNP clumping, the inflation issue for GSMR is largely reduced but there is deflated issue for MRMix. For point estimates, MR-Corr$^2$ is unbiased for all $\rho_{\alpha\gamma}$ in both scenario and has the smallest standard errors while CAUSE is largely biased. Comparing to other alternative methods, GSMR has smaller biasedness and standard deviation for point estimates. The overall pattern is similar when $\rho_{\alpha \gamma} = 0.4$ as shown in {\color{black} Figure S2}. When $\rho_{\alpha \gamma}$ becomes larger, the inflation of type-I error for MR-LDP in Scenario 1 is more severe. This is because MR-LDP  accounts for only IHP but not CHP. We also shows the results from no CHP ($\rho_{\alpha \gamma} = 0$) in Figure S1. When there is no CHP, all methods perform well in Scenario 1 but point estimates from MR-LDP and RAPS are biased in Scenario 2.  %The results for $\rho_{\alpha \gamma} = 0$ and $0.4$ are shown in Figures X and X, respectively, in the Supplementary document. 

\section{Real Data Analysis}\label{real.data}
\subsection{Real validations}
Clearly, when there is no horizontal pleiotropy either correlated or independent, in other words, $\eta_k= 0, \forall k$ , the proposed MR-Corr reduces to the classical MR methods that Eqn.~(\ref{linear0}) holds.  Here, we used real datasets for height as both exposure and outcome to compare the estimates from MR-Corr and MR-Corr$^2$ with those from the other five alternative methods. As the exposure and outcome are the same complex traits, the causal effect $\beta_0$ can be taken as known, i.e., $\beta_0=1$.  Since this type of analysis estimates the causal effects with known effect sizes under no horizontal pleiotropy, we make comparisons among the proposed and alternative methods to see the effectiveness of accounting for weak instrumental variants in LD. 

In this validation study, we treated the summary statistics for height in UK Biobank~\cite{bycroft2018uk} as the screening dataset, and chose the summary statistics for height in male and females in an European
population-based study~\cite{randall2013sex} as exposure and outcome, respectively.  In the analysis, we first selected SNPs under different $p$-value threshold as instrumental variants. 
We next conducted SNP-clumping to obtain near-independent SNPs using PLINK~\cite{purcell2007plink} for MR-Corr, CAUSE, GSMR, MRMix and RAPS.  For both MR-Corr$^2$ and MR-LDP, we conducted the analysis using all SNPs and used the genotype data from UK10K project~\cite{uk10k2015uk10k} merged with 1000 Genome Project Phase 3~\cite{10002015global} as the reference dataset. The estimated causal effects with their corresponding standard errors using seven methods are summarized in Table~\ref{hh.validation}. Clearly, 95\% confidence intervals from MR-Corr, MR-Corr$^2$, MR-LDP, MRMix and RAPS cover the true $\beta_0$ but not for CAUSE and GSMR. Comparing to MR-Corr$^2$, MR-LDP and RAPS, the standard errors from MRMix are very large. On the other hand, using more correlated instrumental variants improves estimation efficiency for MR-Corr$^2$ and MR-LDP as both methods used SNPs in LD. This validation study demonstrates that the proposed methods %MR-Corr, MR-Corr$^2$, MR-LDP and RAPS can all 
perform well when there is no horizontal pleiotropy Eqn.~(\ref{linear0}). Moreover, methods using correlated SNPs, i.e., MR-Corr$^2$ and MR-LDP, are more statistically efficient.

\begin{table}[ht]
	%\centering
	\footnotesize
	\caption{\footnotesize Results for height-height example. We chose 10 different $p$-value thresholds, $p_{\text{sel}}$, to select instrumental variants from $5\times 10^{-8}$ to $1\times 10^{-3}$, and reported the estimated causal effects with their corresponding standard errors for all six methods. Note that SNP clumping was performed to obtain independent SNPs for MR-Corr, CAUSE, GSMR, MRMix and RAPS. }\label{hh.validation}
	\begin{tabular}{llllllll}
		\hline
		%		$p_{\text{sel}}$& MR-Corr\textsuperscript{2}  & MR-LDP & MR-Corr  & CAUSE & GSMR & MRMix & RAPS \\ 
		%		\hline
		%5e-08 &  1.011(0.018) & 1.022(0.020) & 1.012(0.026) & 1.428(0.085) & 0.789(0.021) & 0.805(1.807) & 1.015(0.034) \\ 
		%1e-07 &  1.013(0.019) & 1.024(0.020) & 1.015(0.029) & 1.426(0.085) & 0.784(0.021) & 0.800(6.841) & 1.019(0.035) \\ 
		%5e-07 &  1.013(0.017) & 1.023(0.019) & 1.010(0.028) & 1.422(0.086) & 0.738(0.02) & 0.825(1.71) & 1.017(0.036)\\ 
		%1e-06 &  1.010(0.017) & 1.020(0.019) & 1.005(0.028) & 1.413(0.087) & 0.730(0.019) & 0.780(0.396) & 1.008(0.036) \\ 
		%5e-06 &  1.007(0.018) & 1.020(0.019) & 1.011(0.029) & 1.391(0.089) & 0.709(0.019) & 0.765(0.276) & 1.018(0.039) \\ 
		%1e-05 &  1.005(0.018) & 1.018(0.018) & 0.996(0.030) & 1.346(0.090) & 0.694(0.019) & 0.800(0.639) & 1.004(0.039) \\ 
		%5e-05 &  1.002(0.018) & 1.017(0.018) & 0.992(0.031) & 1.320(0.093) & 0(0) & 0.790(1.573) & 0(0) \\ 
		%1e-04 &  1.001(0.019) & 1.018(0.018) & 0.985(0.028) & 1.317(0.093) & 0(0) & 0.560(0.573) & 0(0) \\ 
		%5e-04 &  0.997(0.018) & 1.017(0.017) & 0.982(0.032) & 1.235(0.099) & 0.562(0.017) & 0.570(0.479) & 1.006(0.053) \\ 
		%1e-03 &  0.993(0.018) & 1.016(0.017) & 0.970(0.032) & 1.178(0.104) & 0.516(0.016) & 0.560(0.234) & 0.995(0.059) \\ 
		$p_{\text{sel}}$& MR-Corr\textsuperscript{2}  & MR-LDP & MR-corr & CAUSE & GSMR & MRMix & RAPS \\ 
		\hline
		5e-08 & {\color{black}1.010(0.019) }& 1.022(0.020) & 1.012(0.026) & 1.428(0.085) & 0.789(0.021) & 0.805(1.807) & 1.015(0.034) \\ 
		1e-07 & {\color{black} 1.011(0.020)} & 1.024(0.020) & 1.015(0.029) & 1.426(0.085) & 0.784(0.021) & 0.800(6.841) & 1.019(0.035) \\ 
		5e-07 & {\color{black} 1.007(0.020)}& 1.023(0.019) & 1.010(0.028) & 1.422(0.086) & 0.738(0.020) & 0.825(1.710) & 1.017(0.036) \\ 
		1e-06 & {\color{black}1.005(0.018) }& 1.020(0.019) & 1.005(0.028) & 1.413(0.087) & 0.730(0.019) & 0.780(0.396) & 1.008(0.036) \\ 
		5e-06 & {\color{black} 1.001(0.019)} & 1.020(0.019) & 1.011(0.029) & 1.391(0.089) & 0.709(0.019) & 0.765(0.276) & 1.018(0.039) \\ 
		1e-05 & {\color{black} 1.001(0.019)}& 1.018(0.018) & 0.996(0.030) & 1.346(0.090) & 0.694(0.019) & 0.800(0.639) & 1.004(0.039) \\ 
		5e-05 & {\color{black} 0.996(0.018)} & 1.017(0.018) & 0.992(0.031) & 1.320(0.093) & 0.649(0.018) & 0.790(1.573) & 0.997(0.042) \\ 
		1e-04 & {\color{black} 0.997(0.019)}& 1.018(0.018) & 0.985(0.028) & 1.317(0.093) & 0.627(0.017) & 0.560(0.573) & 0.991(0.043) \\ 
		5e-04 & {\color{black} 0.989(0.018)} & 1.017(0.017) & 0.982(0.032) & 1.235(0.099) & 0.562(0.017) & 0.570(0.479) & 1.006(0.053) \\ 
		\hline
	\end{tabular}
\end{table}

\subsection{MR analysis for GWAS summary statistics}
\begin{figure} [ht!]
	\begin{subfigure}{.5\textwidth}
		\centering
		% include first image
		\includegraphics[width=.99\linewidth]{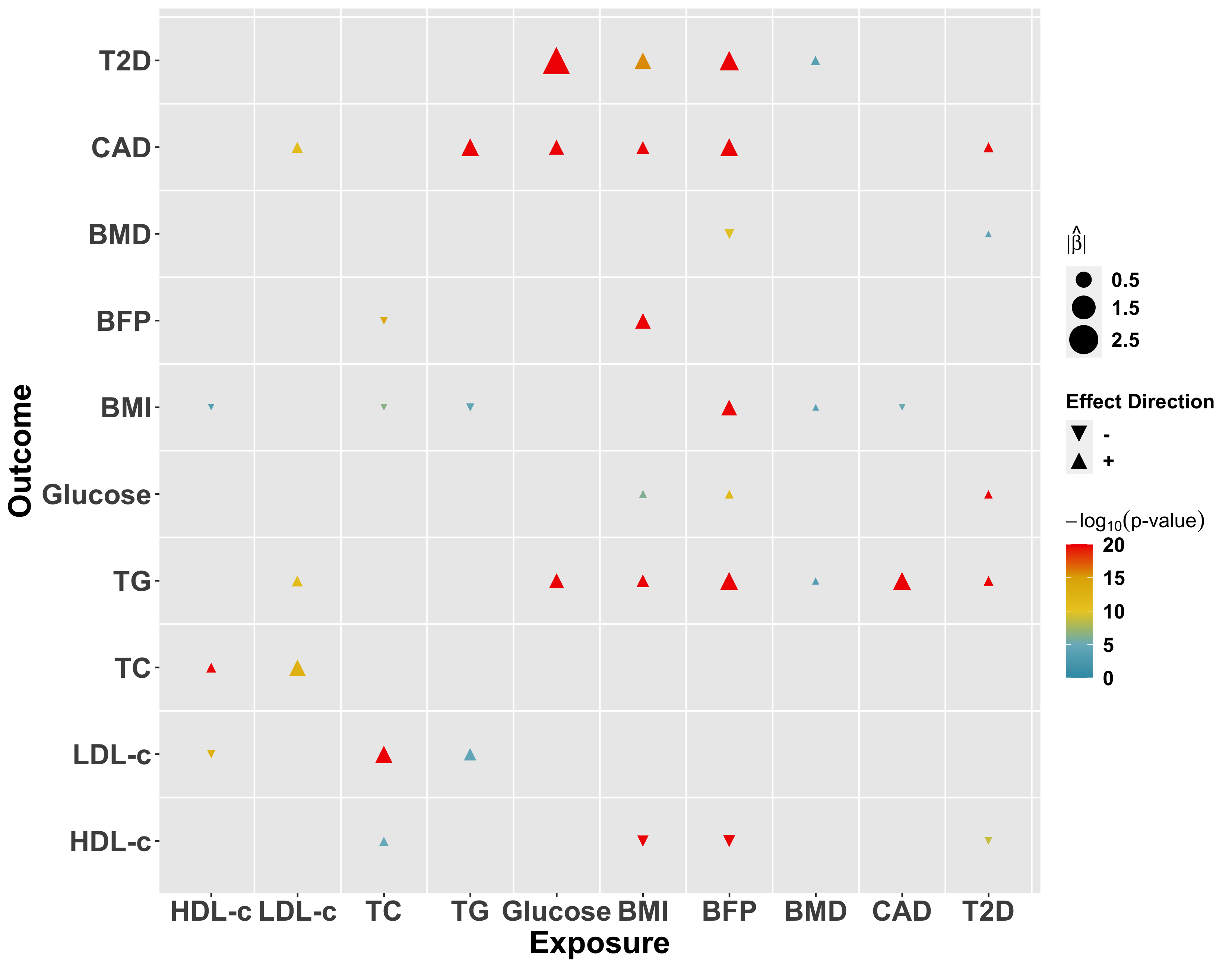}  %39
		\caption{{\color{black}MR-Corr$^2$}}
		\label{fig:sub-mr-corr2}
	\end{subfigure}	
	\begin{subfigure}{.5\textwidth}
		\centering
		% include first image
		\includegraphics[width=.99\linewidth]{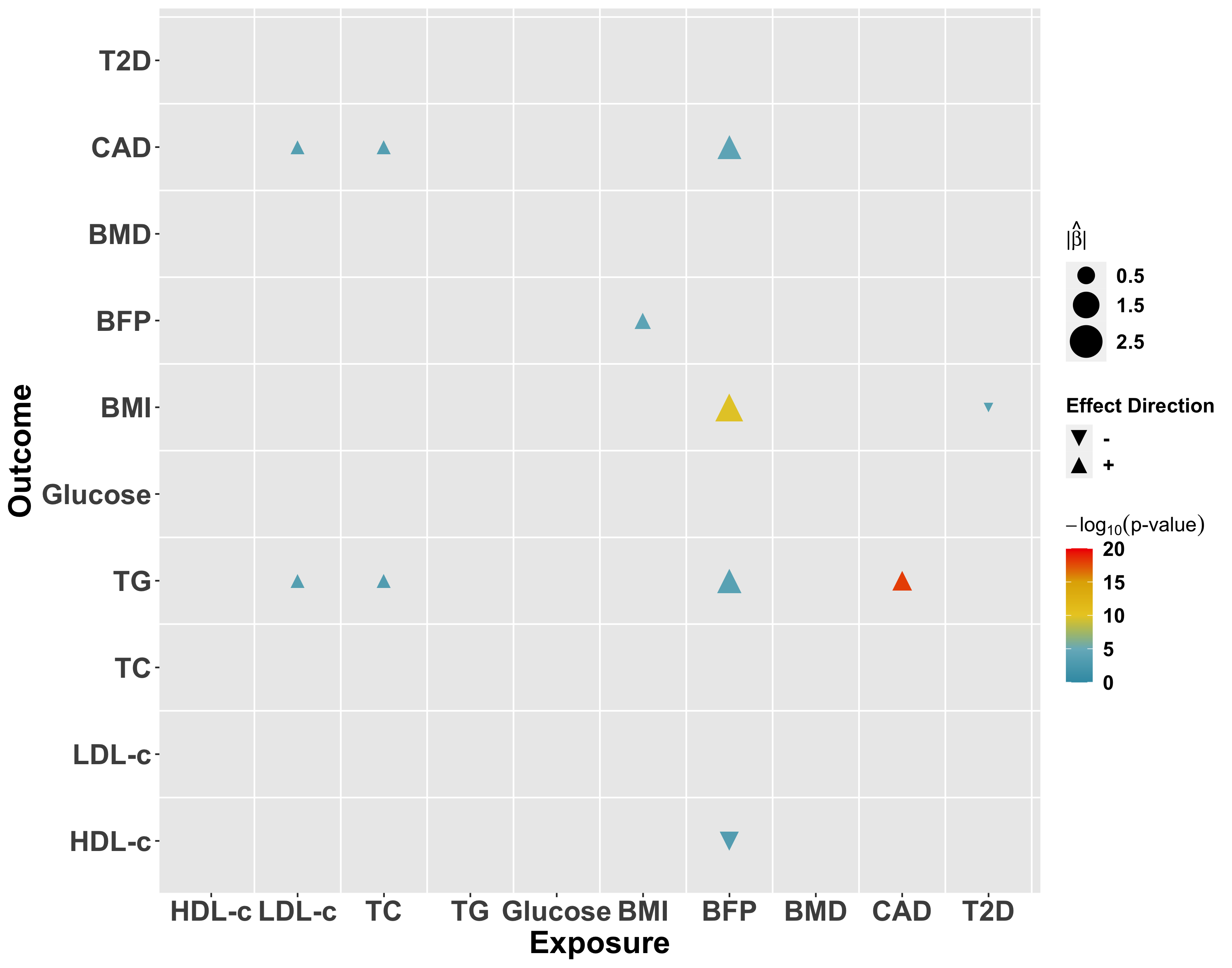}  %10
		\caption{CAUSE}
		\label{fig:CAUSE}
	\end{subfigure}	
	\begin{subfigure}{.5\textwidth}
		\centering
		% include first image
		\includegraphics[width=.99\linewidth]{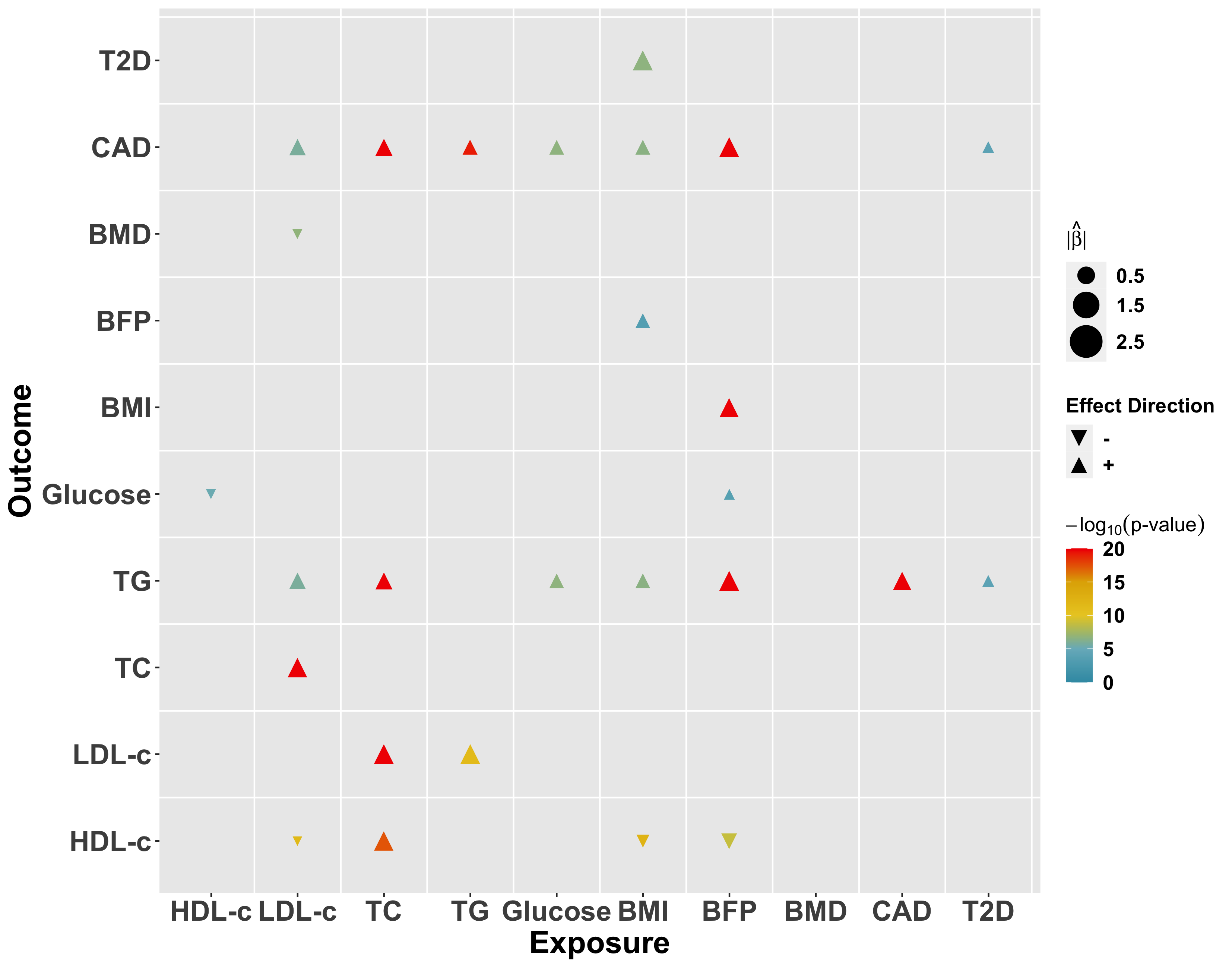}  %25
		\caption{MRMix}
		\label{fig:MRMix}
	\end{subfigure}	
	\begin{subfigure}{.5\textwidth}
		\centering
		% include first image
		\includegraphics[width=.99\linewidth]{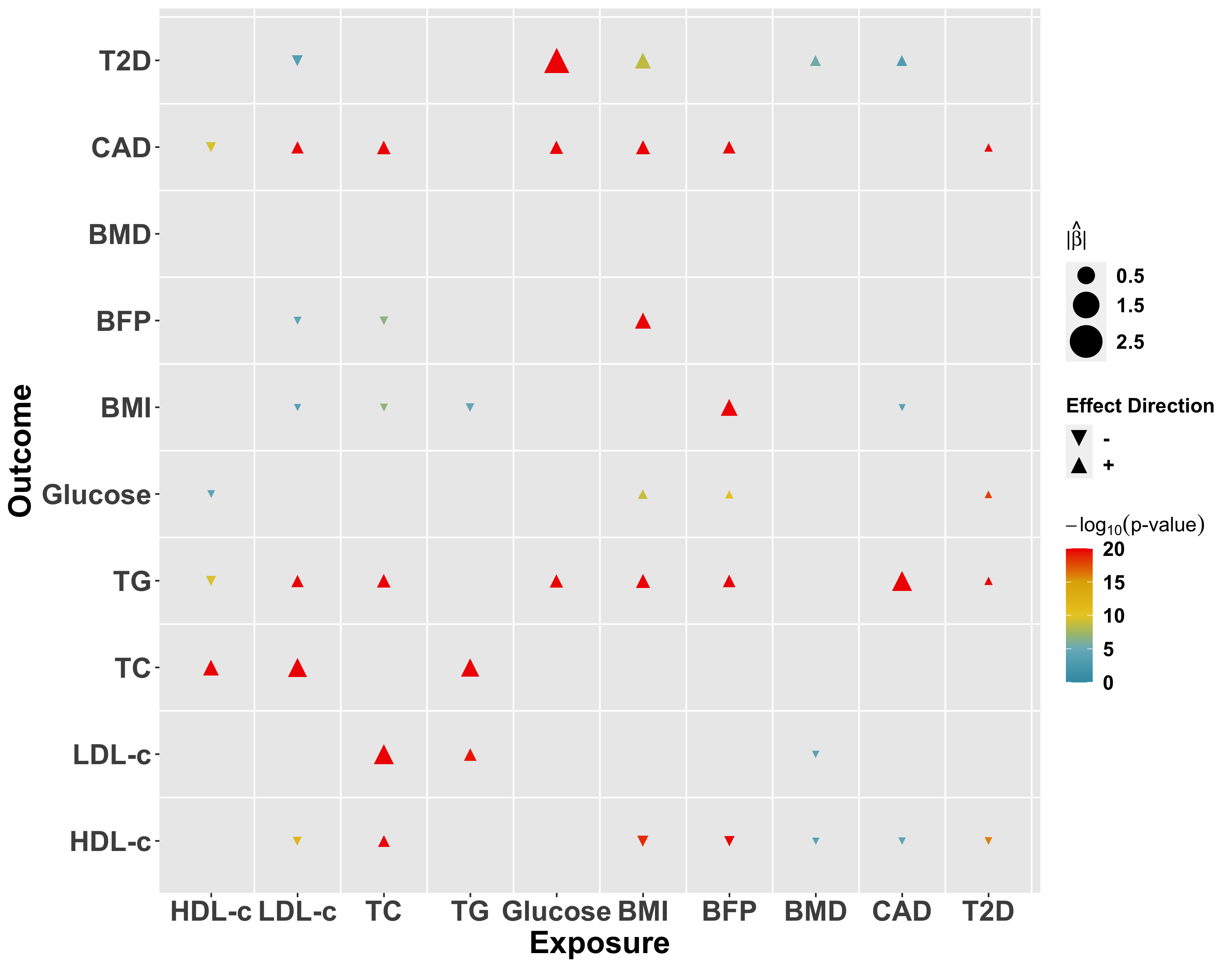}  %39
		\caption{GSMR}
		\label{fig:GSMR}
	\end{subfigure}	
	\begin{subfigure}{.5\textwidth}
		\centering
		% include first image
		\includegraphics[width=.99\linewidth]{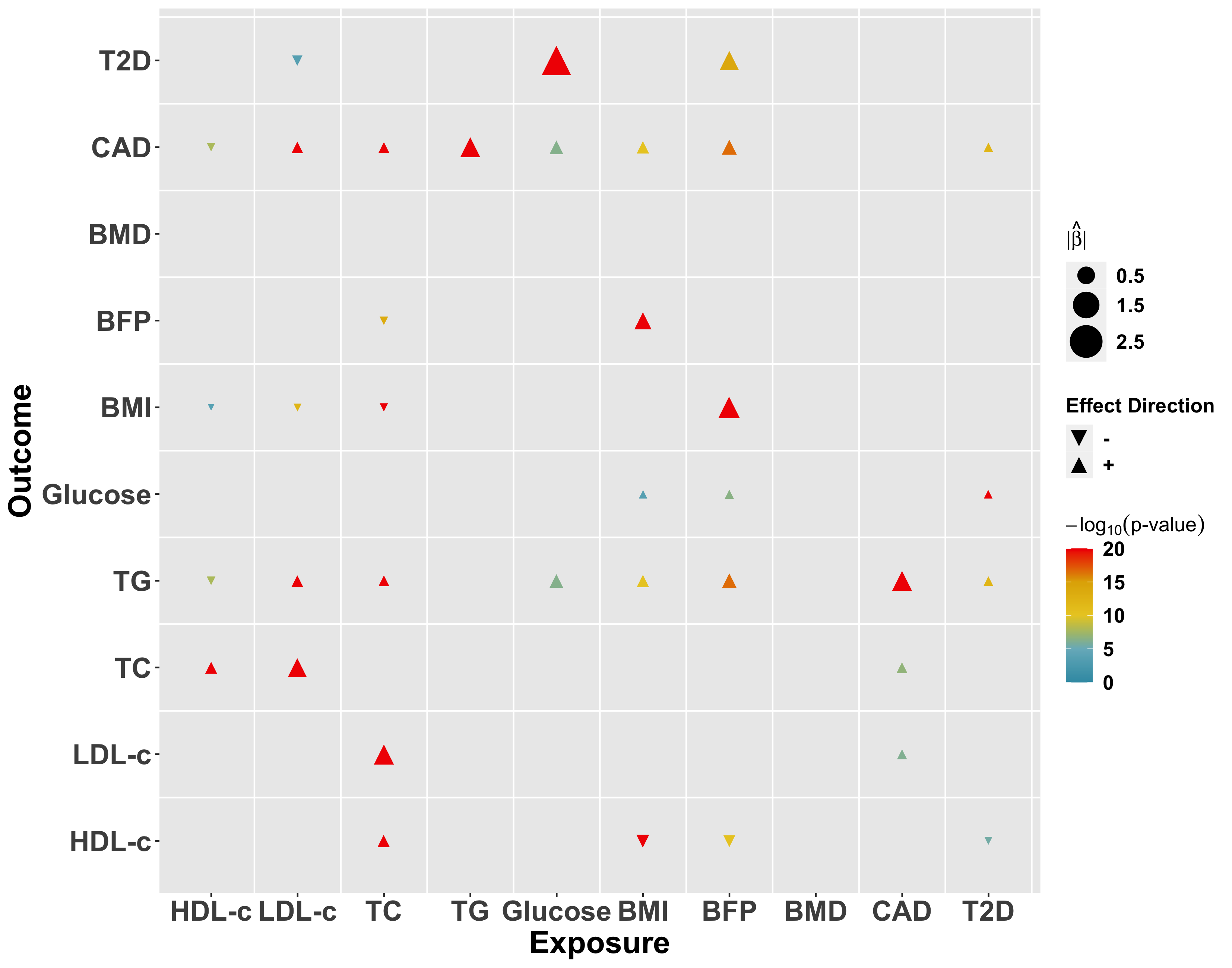}  %37
		\caption{MR-LDP}
		\label{fig:MRLDP}
	\end{subfigure}	
	\begin{subfigure}{.5\textwidth}
		\centering
		% include first image
		\includegraphics[width=.99\linewidth]{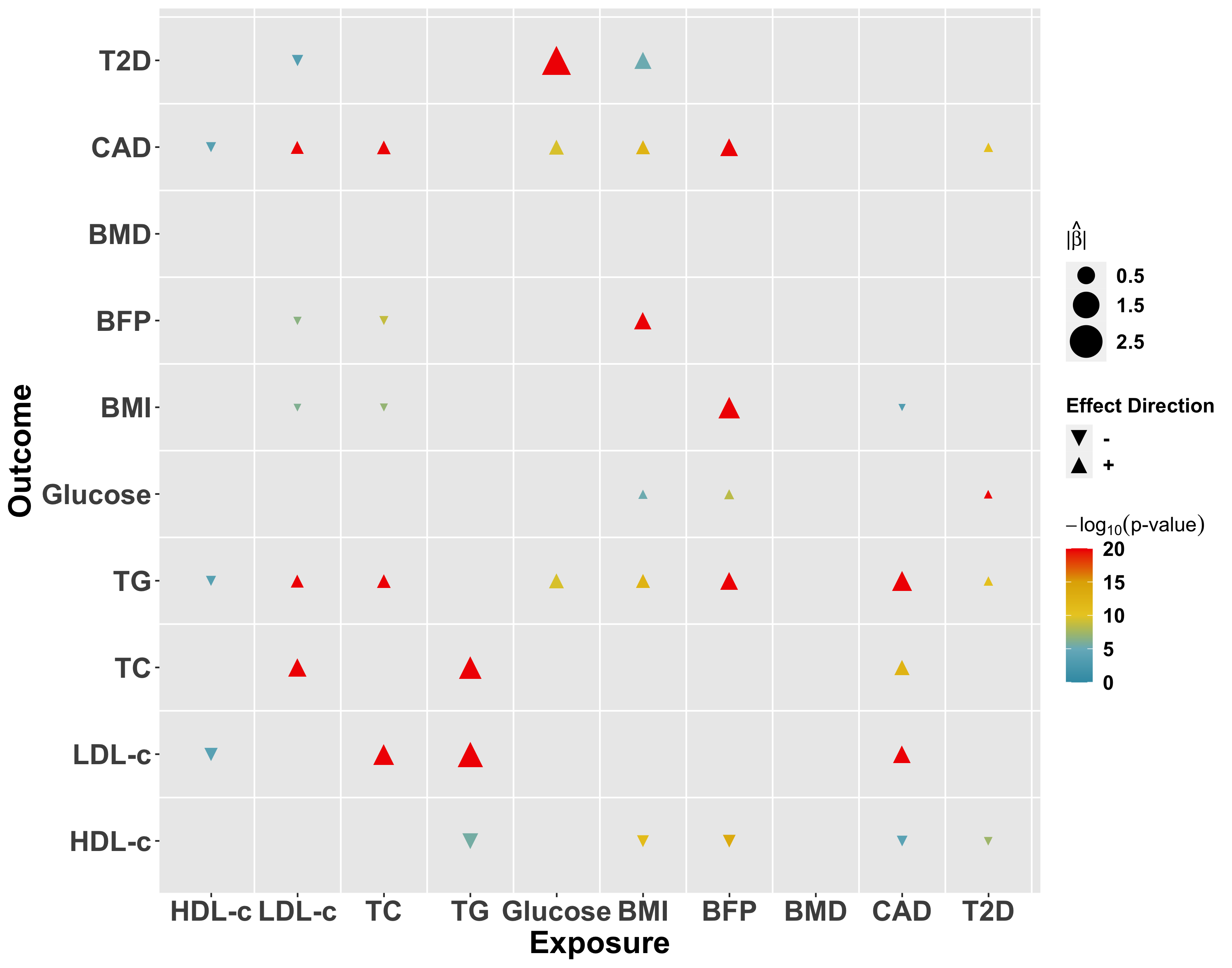}  %37
		\caption{RAPS}
		\label{fig:RAPS}
	\end{subfigure}	
	\caption{Estimated causal effects and their significance between exposure-outcome pairs among nine traits. In this plot, we show the magnitude ($|\hat\beta_0$) and direction (+ or -) of estimated causal effects together with their significance level ($-\log_{10}(p$-value)).}\label{fig:results}
\end{figure}

In the real data analysis, we applied six methods that are used in validation studies to analyze the causal effects between exposure-outcome pairs among two sets of traits. The first set contains {\color{black} ten} traits, including low density lipoprotein cholesterol (LDL-c), high density lipoprotein cholesterol (HDL-c), total cholesterol (TC), triglycerides (TG), glucose, body fat percentage (BFP),  {\color{black}total bone mineral density (BMD)}, body mass index (BMI), coronary artery disease (CAD), type 2 diabetes (T2D). The second set has five psychiatric disorders, including attention deficit hyperactivity disorder (ADHD), major depression disorder (MDD), bipolar disorder (BIP), schizophrenia (SCZ) and autism spectrum disorder (ASD).  In the analysis using CAUSE, MRMix, GSMR and RAPS, we used $r^2=0.1$ to perform SNP clumping by CAUSE default. The results are shown in Figure~\ref{fig:results} for MR-Corr$^2$, CAUSE, MRMix, GSMR,  MR-LDP, and RAPS, respectively.

In the first set, since we conducted MR analysis for $10\times 9= 90$ exposure-outcome pairs, we performed Bonferroni correction for significance level. In Figure~\ref{fig:results}, we identified {\color{black} 39}, 11, 27, 45, 37 and 40 significant causal pairs using MR-Corr$^2$, CAUSE, MRMix, GSMR,  MR-LDP, and RAPS, respectively. %, which is consistent with simulation results that $p$-values for CAUSE and MRMix are deflated but those for GSMR, MR-LDP and RAPS are inflated. 
In details, many MR methods not correcting for CHP identified the significant causal effect of LDL-c on T2D~\cite{pan2020ldl}. But after correcting for CHP, MR-Corr$^2$ did not identify it significantly. One of the driving cause of T2D is insulin resistance~\cite{johnson2013origins}.  %~\cite{shepherd1999glucose, johnson2013origins}. 
And it is known that insulin resistance had profound effects on lipoproteins including LDL-c~\cite{garvey2003effects}. So the variants associated with insulin resistance could affect both T2D and LDL-c level, which is a case of CHP. %{\color{red} more discussion here.}  
As both BMI and BFP measure the level obesity, all six methods identified significant reverse causality between BMI and BFP with effect size close to 1, implying that the causal outcomes for BMI should be similar to those for BFP.  %there should exit reverse causality between them,  Interestingly, all six methods identified significant reverse causality between BMI and BFP.  
As shown in Figure~\ref{fig:results}(a), both BMI and BFP are causally related to T2D based on our {\color{black}MR-Corr$^2$} test. However, all the other methods identified only one of them as the causal risk factor for T2D. It is known that both normal weight individuals with high BFP and overweight individuals with low BFP have higher risk for prediabetes or diabetes~\cite{jo2018informational}. Therefore, only our method can uncover this combined effect of BMI and BFP on T2D. Other than T2D, we also consistently observe this combined effect on CAD as well as several metabolite levels. Besides BMI and BFP, we also identified two reverse causalities of T2D. One of them is T2D and glucose, which is expected because the most important characteristics of T2D is high fasting blood glucose~\cite{defronzo2015type}. The other novel reverse causality exits between T2D and BMD, which is consistent with the previous findings that BMD is increased in T2D patients~\cite{vestergaard2007discrepancies}.
%Only the results from MR-Corr$^2$ show this pattern but not for others. For example,  MR-Corr$^2$ identified significant causal effects of both BMI and BFP on T2D, which is consistent with many completed trials that (more discussions) [cite]. 

Though the causal roles of LDL-c and TG on CAD are confirmed in different clinical trials~\cite{marston2019association}, %\cite{ballantyne1998low, marston2019association}, 
but it is not clear for the role of TC on CAD. After Bonferroni correction, MR-Corr$^2$ did not estimate the casual effect significantly between HDL-c and CAD but identified significantly by GSMR, MR-LDP and RAPS. Our result of MR-Corr$^2$ is consistent with those from clinical trials that use {\it CETP} inhibitor to raise the HDL-c level, but does not result in a lower rate of cardiovascular disease~\cite{lincoff2017evacetrapib}. %~\cite{barter2007effects, schwartz2012effects, lincoff2017evacetrapib}. 
Recent studies show that patients with diabetes are at a substantially increased risk of dying from cardiovascular disease~\cite{fitchett2017heart} while a recent trial shows that the beneficial effects of a sodium-glucose co-transporter 2 ({\it SGLT2}) inhibitor  on prevention of hospitalization for heart failure therefore represent a significant clinical breakthrough~\cite{nassif2019review}. %\cite{wiviott2018design, nassif2019review}. 
The analysis using MR-Corr$^2$, MRMix, GSMR and MR-LDP  confirms these established causal relationships. %{\color{red} discuss glucose and type 2 diabetes reverse causlity.}  The results in Figure~\ref{fig:results} are summarized in Table SX.  %heart failure is now emerging as the most common complication of type 2 diabetes~\cite{nassif2019review}   %one can observe that CAUSE identified the least number of causal pairs, which is consistent with the simulation results that CAUSE has severe problem of $p$-value deflation. On the other hand, GSMR, MR-LDP, and RAPS 

In addition, we also observe several reverse causalities among lipid metabolites, which are TC-HDL-c, TC-LDL-c, and TG-LDL-c. These relationships are expected because they are naturally related with each other. For example, TC is a measure of LDL-c, HDL-c, and other lipid components, and LDL-c level can be calculated from TG~\cite{chen2010modified}.

For the second set of traits, we studied the causal relationships among psychiatric disorders. Since we conducted MR analysis for $5\times 4= 20$ exposure-outcome pairs, we performed Bonferroni correction for significance level. It is well-known genetic correlations among psychiatric disorders are high, primarily due to the fact that common risk variants  for psychiatric disorders are correlated significantly~\cite{anttila2018analysis}. {\color{black}In Figure S4 (Supplementary), we identified 8, 0, 2, 7, 7 and 6 significant causal pairs using MR-Corr$^2$, CAUSE, MRMix, GSMR,  MR-LDP, and RAPS, respectively. }%All methods except CAUSE identify the significance causal effect of SCZ on BIP. Moreover, 
MR-Corr$^2$ identified four psychiatric disorders to be causally affected by MDD while {\color{black} RAPS and MR-LDP identify }three, GSMR identifies two, and {\color{black}  MRMix  identified} one among the four identified by MR-Corr$^2$. Meanwhile, MR-Corr$^2$, {\color{black} MR-LDP}, GSMR and RAPS identified a significant positive causal effect of SCZ on ASD, BIP and MDD, and MR-Corr$^2$ , {\color{black} MR-LDP} and GSMR  identified a significant positive causal effect of ASD on ADHD. Table S2 in the Supplementary document shows the genetic correlations among these psychiatric disorders in~\cite{anttila2018analysis}. A general pattern is that a significant correlation usually implies a causal relationship except for pairs in ADHD-BIP and ADHD-SCZ. These causal relationships may be attributable to the existence of a large number of medical comorbidities in psychiatric disorders~\cite{plana2019exploring}.    %As shown in Table~\ref{gcor}, %Since \cite{anttila2018analysis} reports that the genetic correlation is 0.521 ($p$-value = $2.18E-20$) between ADHD and MDD, 0.351 ($p$-value = $2.75E-28$) between BIP and MDD, 0.155 ($p$-value = $0.0099$) between ASD and MDD, and 0.338 ($p$-value = $5.45E-33$) between SCZ and MDD, it is very likely there are causal relationships between these four psychiatric disorders and MDD. Without applying MR methods, we cannot identify the cause and the outcome in a pair.
{\color{black}The detailed results for causal inference and Venn plot for psychiatric disorders are shown in Table S3 and Figure S5 in the Supplementary document.}
%The results from the second set among psychiatric disorders are shown in the supplementary document Section S5.2. 

%	\includegraphics[width=0.6\columnwidth]{Figure/MRcorr21e4June17.png}\\
%		\includegraphics[width=0.6\columnwidth]{Figure/CAUSE1e4June17.png} \\
%		\includegraphics[width=0.6\columnwidth]{Figure/MRLDP1e4June17.png}\\
%		\includegraphics[width=0.6\columnwidth]{Figure/MRMix1e4June17.png} \\

%VennplotJul07

%\begin{figure}
%	\centering
%	\includegraphics[width=.8\linewidth]{Figure/VennplotJul07.png}  %37
%	\caption{Venn diagram for the significant pairs identified among six methods.}\label{fig:venn}
%\end{figure}
%\clearpage
\section{Discussion}
In this study, we proposed  MR-Corr$^2$ to account for CHP in the presence of correlated genetic variants. Mitigating the impact of CHP is important in MR/IV methods.  We proposed a MLR strategy that first decomposes the direct effects into two parts, linear and orthogonal, then reparameterizes the relationship between $\bfgamma$ and $\bfGamma$. In this way, the impact of CHP can be mitigated using model with variable selection, e.g., a spike-slab prior is used here. To incorporate correlated SNPs, we %extended MR-Corr to MR-Corr$^2$ using
use an approximated distribution for summary statistics. Our validation studies which include both a real validation and simulations demonstrate that MR-Corr$^2$ controls type-I error at its nominal level and estimates the causal effects unbiasedly. %and  $p$-value of MR-Corr$^2$ is little bit conservative that is primarily due to the application of approximated summary statistics distribution for correlated SNPs. 
In real data analysis, MR-Corr$^2$ did not identify the contradictory causal relationship between HDL-c and CAD, but identified multiple health outcomes, i.e., T2D, CAD, TG, and HDL-c, affected by obese risk factors (BMI and BFP) causally.  

The proposed MLR strategy provides an effective and systematic framework to mitigate the impact of CHP in MR/IV methods. By connecting MLR strategy with CAUSE, it offers a unique perspective from correlated pleiotropy that is genetic variants affect unobserved confounders as shown in Figure~\ref{fig:mech2}. When there exist both correlated pleiotropy and IHP, an ideal model should allow linear relationship~(\ref{linear0}) deviates with IHP. Specifically, other than using Eqn.~(\ref{relat}), we now can model an additional IHP, $\theta_k$, as 
\be\label{relat2}
\Gamma_k =  \left\{
\begin{aligned}
	&\beta_0 \gamma_k + \theta_k ,   &&\wt \alpha_k  = 0  \\
	&\beta_1 \gamma_k  + \theta_k+  \wt \alpha_k,  & &\wt \alpha_k \neq 0.
\end{aligned}
\right.
\ee
In this way, we will be able to handle both correlated pleiotropy as well as {\color{black} independent} horizontal pleiotropy in the sense of CAUSE. 

The other limitation of the proposed methods is that they do not account for sample overlap between SNP-exposure and SNP-outcome. For example, the samples are largely overlapped for lipid traits, i.e., HDL-c, LDL-c, TC and TC. Without correcting for these overlapped samples, the hypothesis test for causal effects between lipid traits tend to be inflated. We plan to address these challenges in future work.

% \newpage
% \bibliographystyle{plain}
\bibliographystyle{abbrv}
\bibliography{MRLDCP}

\end{document}